\documentclass[journal, a4paper]{IEEEtran}
\usepackage[T1]{fontenc}
\usepackage[latin9]{inputenc}
\usepackage{color}
\usepackage{amsmath}
\usepackage{amsthm}
\usepackage{amssymb}
\usepackage{esint}
\usepackage{verbatim}
\usepackage{graphicx}
\usepackage{stfloats}
\usepackage{subfigure}
\usepackage{dsfont}
\usepackage{bm}
\usepackage{balance}
\usepackage{multicol}
\usepackage{multirow}
\usepackage{tabularray}
\usepackage{makecell}
\usepackage{algorithm}
\usepackage{algorithmic}
\usepackage{mathtools}
\usepackage{array}
\usepackage[numbers,sort&compress]{natbib}

\makeatletter

\newtheorem{assp}{Assumption}

\definecolor{sblue}{RGB}{0,0,0}
\begin{document}
\title{
 Data-Importance-Aware Power Allocation for Adaptive {Semantic} Communication in Computer Vision Applications\\
}
\author{Chunmei Xu,~\IEEEmembership{Member,~IEEE}, Yi Ma,~\IEEEmembership{Senior Member,~IEEE}, Rahim Tafazolli,~\IEEEmembership{Fellow,~IEEE},  \\ Jiangzhou Wang,~\IEEEmembership{Fellow,~IEEE}

\thanks{This work was partially funded by the UKRI-EPSRC under Intelligent Spectrum Innovation (ICON) programme (APP55159).}

\thanks{C.~Xu, Y. Ma and, R. Tafazolli are with 5GIC \& 6GIC,
Institute for Communication Systems (ICS), University of Surrey, Guildford,
U.K. (emails:\{chunmei.xu; y.ma; r.tafazolli\}@surrey.ac.uk). (Corresponding authors:  \emph{Y.~Ma, C.~Xu})}

{\color{blue}\thanks{J.~Wang  is with the School of Engineering, University of Kent, CT2 7NT
Canterbury, U.K. (e-mail: j.z.wang@kent.ac.uk).}}

}

\maketitle
\begin{abstract}

Life-transformative applications such as immersive extended reality are revolutionizing wireless communications and computer vision (CV). This paper presents a novel framework for importance-aware adaptive data transmissions, designed specifically for real-time CV applications where task-specific fidelity is critical. A novel importance-weighted mean square error (IMSE) metric is introduced as a task-oriented measure of reconstruction quality, considering sub-pixel-level importance (SP-I) and semantic segment-level importance (SS-I) models. To minimize IMSE under total power constraints,  data-importance-aware waterfilling approaches are proposed to optimally allocate transmission power according to data importance and channel conditions, prioritizing  sub-streams with high importance.  Simulation results demonstrate that the proposed approaches significantly outperform margin-adaptive waterfilling and equal power allocation strategies. The data partitioning that combines both SP-I and SS-I models is shown to achieve the most significant improvements, with  normalized IMSE gains exceeding $7\,$dB and $10\,$dB over the baselines at high SNRs ($>10\,$dB). These substantial gains highlight the potential of the proposed framework to enhance data efficiency and robustness in real-time CV applications, especially in bandwidth-limited and resource-constrained environments. 
\end{abstract}


\section{Introduction}
Life-transformative applications such as immersive extended reality (XR), telemedicine, autonomous systems, digital twins, and the metaverse are driving rapid advancements in wireless communications and computer vision (CV) \cite{giordani2020toward, wang2023road, semeraro2021digital, wang2023survey}. These applications require unprecedented network performance in terms of data rates, latency, and reliability to deliver real-time, interactive experiences that could redefine healthcare, industrial automation, and personal connectivity. 
Achieving such capabilities will push the boundaries of both communication networks and CV technologies.

For future networks (namely sixth-generation $6$G), this means supporting ultra-high data rates (up to $1$ Tbps), sub-millisecond latency (under 1 ms), ultra reliability (99.99999\%), and  cm-level sensing accuracy (under 1 cm). 
Real-time immersive applications like XR cannot afford delays from {\color{sblue}complex}  compression processes, as these would introduce latency that could cause motion sickness or pose risks in telesurgery settings \cite{giordani2020toward}. Consequently, uncompressed data transmission becomes essential, challenging traditional communication paradigms and driving the need for innovative network architectures that can efficiently handle massive uncompressed data streams while maintaining strict performance requirements. CV, meanwhile, is central to these transformative applications, empowering machines to  perceive, process, and understand  visual information from the digital world through sophisticated algorithms and deep learning techniques \cite{szeliski2022computer,voulodimos2018deep}. CV enables features essential for these revolutionary applications: from real-time object tracking \cite{wang2020towards}, facial recognition \cite{hu2015face}, and gesture detection in XR applications, to accurate virtual-to-physical mapping in digital twins and metaverse applications \cite{moya2022digital}. 

However, CV and telecommunications have historically developed along distinct lines, leading to a fundamental divergence in their performance metrics and optimization objectives. CV primarily focuses on task-specific performance \cite{wang2004image,rezatofighi2019generalized},  including mean squared error (MSE) and peak signal-to-noise
ratio  for image restoration, precision and recall
for object detection, accuracy for classification tasks,
and intersection over union  for segmentation. These
metrics reflect the effectiveness of CV algorithms in
understanding and processing visual information. In contrast,
telecommunications prioritizes transmission-oriented metrics
such as data rate, latency, bit error rate (BER), and spectrum
efficiency, which characterize the efficiency and reliability of
data transmission through wireless channels \cite{thomas2006elements}. This divergence
in performance evaluation reveals a critical limitation that
conventional communication systems, often optimized for data
fidelity rather than the specific needs of CV applications,  may lead to an inefficient use of radio resources. For instance, perfectly reconstructing background pixels in facial recognition may consume valuable radio resources without improving task performance. 
Task-oriented semantic communications (SemCom) has emerged as a promising solution by transmitting only the essential ``meaning'' relevant to a task, thus improving resource efficiency and aligning with the requirements of CV  applications \cite{carnap1952outline, gunduz2022beyond}.  

Task-oriented SemCom,  often grounded in joint source-channel coding (JSCC) and deep learning, is designed to focus on the representation and transmission of semantic content critical to a CV task (such as facial features in recognition) rather than transmitting all image details.  End-to-end neural network architectures are  typically employed to learn JSCC that encode and decode the source with semantic equivalence. Deep JSCC has the potential to improve both communication efficiency  and task-specific performance \cite{xie2021deep, weng2021semantic, erdemir2023generative, tong2024multimodal}. However, several practical challenges remain, requiring analog modulation and often lacking generalization across diverse tasks. To address these limitations, recent research has proposed a promising solution through the employment of pre-trained foundation models as semantic encoder for feature extraction and decoders for source regeneration   \cite{xu2024semantic, xu2024generative}.  These foundation models offer enhanced system compatibility with existing communication systems, and broaden applicability by training on diverse CV scenarios \cite{bommasani2021opportunities}. 

Despite these advancements, existing SemCom approaches and traditional transmission methods overlook the varying importance of visual information within CV tasks. This content importance stands as the key characteristic for emerging real-time CV applications. Existing transmission strategies  primarily focus on maximizing spectrum efficiency and enhancing reliability through physical layer innovations \cite{gao2016energy, Chaccour2022THz, Larsson2014mmMIMO,Liu2024ELAA, Tal2013Polar}, where the data is treated with equal importance \cite{ guiacsu1971weighted}. This misalignment between data with varying levels of importance and existing transmission strategies inevitably leads to waste of radio resources and thereby degrades system performance.  To address this, a paradigm shift in communication strategies is essential: one that captures the hierarchical importance of visual information and aligns resource allocation with task-specific needs. Such a paradigm shift requires fundamentally rethinking how to evaluate and optimize wireless transmission for CV applications. This brings forth two critical research questions: {\it 1)} How to model a novel metric that reflects the interdependence of CV task requirements, data importance, and telecommunication performance? {\it 2)} How can radio resources be allocated efficiently based on this new metric?

This paper aims to address these questions, with the main contribution summarized as follows:
\begin{itemize}
\item A novel importance-aware data transmission framework is proposed, where data is partitioned into sub-streams with varying levels of importance based on their contribution to specific CV tasks. Data importance is characterized through  bit positions within pixels and semantic relevance within visual segments. Building upon these importance models, three data partitioning criteria are developed: two based on individual models and one combining both models.

\item A novel metric termed importance-weighted mean square error (IMSE) is introduced based on the developed importance models, with three specific expressions derived for the respective importance-aware data partition criteria. This metric provides a task-oriented measure of reconstruction quality, capturing both the task-specific significance of visual information and the interdependence between CV and communication performance.

\item Data-importance-aware waterfilling approaches are developed  under the three proposed importance-aware data partition criteria to minimize IMSEs subject to total power constraints. The optimal power allocation  adapt to both data importance and channel conditions, allocating a lager share of power resource to the sub-streams with higher importance but not necessarily with exceptionally good channels. The data-importance-aware waterfilling gain becomes more pronounced when data importance exhibits high variations. 

\item Simulation results demonstrate the superior performance of the proposed approach to margin-adaptive (MA) waterfilling and equal power allocation methods, with the most significant gains in jointly considering the sub-pixel-level and segment-level importance.  At high SNRs ($ > 10$ dB), the achieved normalized IMSE gains are more than $7\,$dB and $10\,$dB. Additionally, to reach a satisfactory normalized IMSE performance ($-26\, \mathrm{dB}$), the proposed method reduces the required SNR by  $5\,\mathrm{dB}$ and $10\,\mathrm{dB}$ respectively compared to the baselines. These significant improvements highlight the framework's potential to enhance data efficiency and robustness in real-time CV applications, particularly in bandwidth-limited and resource-constrained environments.

\end{itemize}

{\color{black}The rest of the paper is organized as follows. Section II presents the considered data-importance-aware communication model, including wireless transmission model and importance-aware data partitioning. Section III introduces a new task-oriented measure, termed IMSE, and derives respective expressions under different data partitioning criteria. The power allocation problems and the data-importance-aware waterfilling strategies are provided  in Section IV. Extensive simulation and the conclusion are given in Sections V and VI, respectively.}

\section{Data-Importance-Aware Communication Model\label{sec: system model}}

\begin{figure}[tbp]
	\vspace{1em}
	\centering
	\includegraphics[width=1\columnwidth]{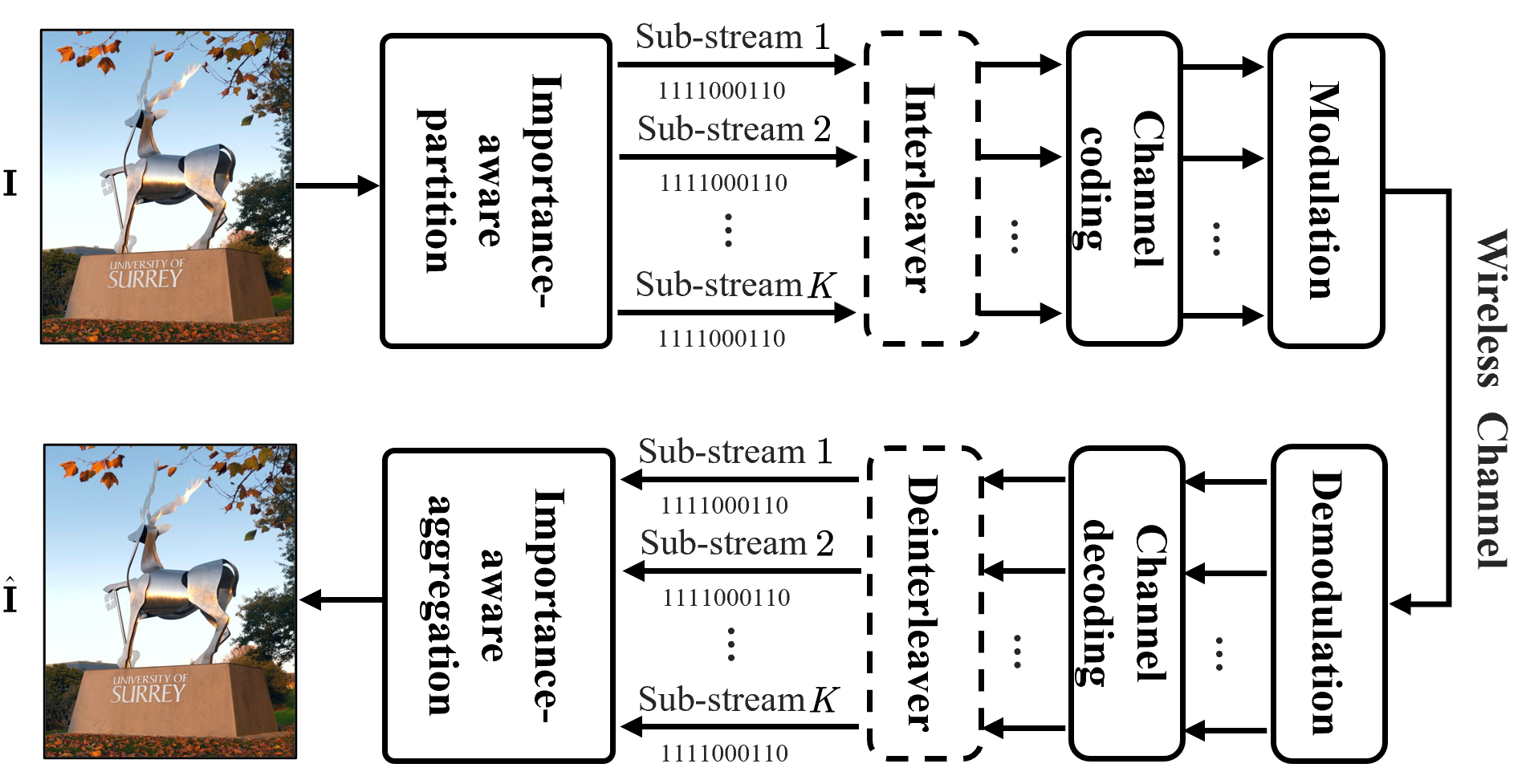}
	
	\caption{The proposed data-importance-aware communication model.}
	\label{fig:system model}
\end{figure}

 Fig. \ref{fig:system model} illustrates the point-to-point model of data-importance-aware communication in real-time CV  applications. 
 The information source is a high-definite image represented as a pixel matrix $\mathbf{I}$ of size $H \times W$, which is uncompressed due to unprecedented latency requirements.
 Each pixel contains multiple color channels, with each channel's pixel values represented by $B$ bits. For the sake of presentation clarity, this paper focuses on a single color channel, as the principles apply to all color channels.
 
 \subsection{Wireless Transmission Model}\label{2a}
 
 Prior to transmission, the pixel matrix $\mathbf{I}$ is partitioned into $K$ bit streams, each with a different level of data importance (see Sec. \ref{sec2b} for details.)
 Each bit stream is individually passed through a random bit interleaver and then fed into a channel encoder {\color{sblue}with a coding rate of $R$}.
 After $M$-quadrature amplitude modulation (QAM), the information-bearing symbol streams, denoted as $\mathbf{x}_k, _{\forall k\in[1, K]}$, are transmitted through their corresponding sub-channels $h_k$ with transmission power $p_k$. 
 The symbol streams received at the receiver, denoted as $\mathbf{y}_k$, are expressed as:
 \begin{equation}\label{eq01}
 	\mathbf{y}_{k} = h_{k}\sqrt{p_{k}}\mathbf{x}_{k} + \mathbf{v}_{k},~ k=1,\dots,K,
 \end{equation}
 where $h_{k}$ is flat block fading following the Rayleigh distribution with the variance of $\sigma^2_{c}$, i.e., $h_{k}\sim \mathcal {CN}(0,\sigma^2_{c})$, and $\mathbf{v}_k$ is the additive white Gaussian noise (AWGN) with zero mean and varience of $\sigma^2$. 
 Given that $\mathbb{E}(\mathbf{x}_k^\mathrm{H}\mathbf{x}_k)=L_k$ (power normalization), the signal-to-noise ratio (SNR) for the $k$-th stream with length $L_k$ is given by
 \begin{equation}\label{eq02}
 	\mathrm{snr}_{k} = \frac{p_{k}\vert h_{k} \vert^2}{\sigma^2},
 \end{equation}
 where $\mathbb{E}(\cdot)$ stands for the expectation, and $(\cdot)^\mathrm{H}$ for the Hermitian. 
 
 Given that each sub-stream is independently coded and decoded, we use the following bit-error-probability (BEP) model to represent the BER \cite{goldsmith2005wireless}:
 \begin{equation}\label{eq:BER_SNR}
 	\mathcal P^\mathrm{e}_{k} = \alpha\exp\left(\beta\mathrm{snr}_{k}\right).
 \end{equation}Here, $\alpha> 0$ and $\beta< 0$ are  parameters determined by the adopted channel coding and modulation schemes, which can be obtained through data fitting (See Appendix \ref{appendixA}). 
 After undergoing a reverse process at the receiver, the pixel matrix is reconstructed, which is then used for CV-specific tasks.
 
 \subsection{Importance-Aware Data Partitioning}\label{sec2b}
The data importance can be modeled based on bit positions within pixels and semantic relevance within visual segments, which are elaborated as follow.
 \begin{enumerate}
 	\item[1).] {\bf Sub-pixel-level importance (SP-I):}  Denote $\mathbf{I}(i,j)$ as the $(i,j)$-th entry of $\mathbf{I}$. It can be represented in polynomial form as:
 {\color{sblue}	\begin{equation}\label{eq03}
 		\mathbf{I}(i,j) = \sum_{b=1}^{B} \mathcal{B}_{i,j}^b \cdot 2^{b-1},
 	\end{equation}
 	where $\mathcal{B}_{i,j}^b\in\{0, 1\}$ represents the $b$-th bit of $\mathbf{I}(i,j)$.
 	An error in the $b$-th bit (where $\hat{\mathcal{B}}_{i,j}^b \neq \mathcal{B}_{i,j}^b$) introduces an error magnitude of $2^{2(b-1)}$, highlighting that bit position within a pixel significantly affects the error magnitude.} Consequently, errors in higher-order bits can severely impact CV task performance, underscoring the need to prioritize accurate transmission for more critical bits. We quantify the importance of the $b$-th bit by its potential error magnitude, modeled as $\gamma_b = 2^{2(b-1)}$. {\color{sblue}This model is fundamentally connected to the MSE metric that serves  as a standard distortion metric across numerous CV applications. Therefore, the SP-I model inherently provides generalizability to any CV task that employs MSE or its derivatives as performance metrics.}
 	
 	\item[2).] {\bf Semantic segment-level importance (SS-I):} The source image can be semantically divided into $S$ segments using state-of-the-art segmentation models, such as the segment anything model (SAM) \cite{kirillov2023segment}. Each visual segment exhibits varying semantic relevance to the specific CV task; for instance, background segments generally contain less task-critical information than object segments. We model the importance of the $s$-th segment as a non-negative value $\gamma_s \ge 0$, representing its relevance to the CV task, with $\sum_{s=1}^S \gamma_s = 1$. {\color{sblue}  It is important to note that the specific values of  $\gamma_s$ are task-dependent and vary across different CV applications. While this model accommodates these varying importance weights, the determination of  $\gamma_s$ values for specific CV tasks remains an open research topic beyond the scope of this work.}
 \end{enumerate} 
 
Based on BP-I, SS-I, or their combination, three importance-aware data partitioning criteria are developed to partition the  pixel matrix $\mathbf{I}$ into $K$ sub-streams.  
 \begin{enumerate}
	\item[1).] {\bf SP-I partitioning:} By using the SP-I model,  all bits located at the same position across  pixels are grouped into one sub-stream, creating $K=B$ sub-streams.    
	
	\item[2).] {\bf SS-I partitioning:}  By using the SS-I model, all bits from pixels within the same semantic segment are grouped into one sub-stream, yielding $K=S$ sub-streams.   

	\item[3).] {\bf SP-SS-I partitioning:}  By combining SP-I ad SS-I models,  bits that share both the same position and semantic segment are grouped into one sub-stream, producing  { $K=(S)(B)$} sub-streams. 
\end{enumerate}

\section{Importance-weighted Mean Square Error}
Conventionally, the error in source reconstruction is measured using the scaled Euclidean norm:
\begin{equation}\label{eq04}
	\epsilon=\frac{1}{I}\|\hat{\mathbf{I}}-\mathbf{I}\|^2{\color{sblue},}
\end{equation}
where $\hat{\mathbf{I}}$ is the reconstructed version of $\mathbf{I}$. $I=(H)(W)$ is the number of pixels of the source image $\mathbf{I}$. 
For a sufficiently large image (e.g., as $I\rightarrow\infty$), the error $\epsilon$ approximates the MSE.

Following the principle of SP-I partitioning (specifically as outlined in \eqref{eq03}), the pixel matrix $\mathbf{I}$ can be represented as:
\begin{equation}\label{eq05}
	\mathbf{I}=\sum_{b=1}^{B} \mathbf{B}_b \cdot 2^{b-1}{\color{sblue},}
\end{equation}
where $\mathbf{B}_b$ is a binary matrix with {\color{sblue}$\mathbf{B}_b(i,j)=\mathcal{B}_{i,j}^b$} in \eqref{eq03}.
Plugging \eqref{eq05} into \eqref{eq04} results in
{\color{sblue}\begin{IEEEeqnarray}{ll}
	\epsilon
	=\frac{1}{I}\Big\|\sum_{b=1}^{B}\sqrt{\gamma_b}(\hat{\mathbf{B}}_b-\mathbf{B}_b)\Big\|^2\label{eq07},
\end{IEEEeqnarray}}where $\hat{\mathbf{B}}_b$ is a binary matrix of $\hat{\mathbf{I}}$. This MSE representation, however, is not well-suited to the optimization task that will be addressed in Sec. \ref{sec4}. 
To address this, we introduce the following assumption.

\begin{assp}\label{assp1}
	At most one bit out of $B$ bits within a pixel is incorrectly reconstructed due to communication errors. 
\end{assp} 
Under this assumption, we obtain:
\begin{equation}\label{eq08}
	(\hat{\mathbf{B}}_{b1}-\mathbf{B}_{b1})\odot(\hat{\mathbf{B}}_{b2}-\mathbf{B}_{b2})=\mathbf{0}, ~\forall b1\neq b2,
\end{equation}
allowing us to simplify \eqref{eq07} as:
\begin{equation}\label{eq09}
	\epsilon=\frac{1}{I}\sum_{b=1}^{B}\gamma_b\|\hat{\mathbf{B}}_b-\mathbf{B}_b\|^2,
\end{equation}
where $\odot$ denotes the matrix Hadamard product.

Building further on the principle of SS-I partitioning, $\mathbf{B}_b$ is decomposed into $S$ sub-matrices, 
denoted by $\mathbf{B}_b^{(s)}$,
each corresponding to a distinct semantic segment. 
Then, \eqref{eq09} can be further expressed as

\begin{align}\label{eq10}
	\epsilon =\sum_{b=1}^{B}\gamma_b\sum_{s=1}^{S}\frac{\|\hat{\mathbf{B}}_b^{(s)}-\mathbf{B}_b^{(s)}\|^2 }{I},
\end{align}
Note that this MSE model does not capture the varying importance of semantic segments, which is crucial for CV-specific tasks. To address this limitation, we introduce a new task-oriented metric, termed IMSE, as:
{\color{black}\begin{IEEEeqnarray}{ll}\label{eq11}
		\mathrm{imse} =\sum_{b=1}^{B}\gamma_b\sum_{s=1}^{S}\gamma_s \frac{\|\hat{\mathbf{B}}_b^{(s)}-\mathbf{B}_b^{(s)}\|^2}{I_{b,s}} ~\mathrm{s.t.}  \sum_{s=1}^S \gamma_s = 1,
\end{IEEEeqnarray}}where $I_{b,s}$ denotes the number of bits within the sub-matrix $\mathbf{B}_b^{(s)}$. The SP-I and SS-I are reflected by $\gamma_b$ and $\gamma_s$ as already discussed in Sec. \ref{sec2b}.

Our resource allocation strategy (in Sec. \ref{sec4}) then seeks to minimize the IMSE through optimum multi-sub-stream (or equivalently multi-sub-channel) power allocation. The IMSE expression in \eqref{eq11} is however not ready to use as it lacks an explicit relationship to the signal power. To address this, we will reformulate the IMSE to incorporate power dependencies, enabling a more effective optimization of power allocation in accordance with the data importance of each sub-stream.  Let $e_{b,s}$ be the reconstruction error of $\mathbf{B}_b^{s}$, which is given by:
\begin{equation}\label{eq12}
	e_{b,s}=\frac{1}{I_{b,s}}\|\hat{\mathbf{B}}_b^{(s)}-\mathbf{B}_b^{(s)}\|^2.
\end{equation}

The reformulated IMSEs  under the proposed SP-I, SS-I and SP-SS-I partitioning criteria are {\color{sblue}provided as follows. For the SP-I model, $\mathbf{B}_b$ forms the transmitted sub-stream $\mathbf{b}_b$ with uniform error $e_{b,s}$ across all segments. With the BER denoted as $\mathcal P^\mathrm{e}_{b}$, we have $e_{b,s} =\mathcal P^\mathrm{e}_{b},  \forall s=1,\dots, S$. After some tidy-up work,  the IMSE form in \eqref{eq11} can be represented as: 
\begin{equation}\label{eq:imse1}
	\mathrm{imse}(p_{b}) 
	=\sum_{b=1}^{B}\gamma_b\alpha \exp\left(\beta\frac{p_{b}\vert h_{b} \vert^2}{\sigma^2}\right),
\end{equation}where $p_b$ is the power allocated to each symbol of the $b$-th sub-stream over the sub-channel $h_b$.  

For the SS-I partitioning, $[\mathbf{B}_1^{(s)}, \dots, \mathbf{B}_B^{(s)}]$ forms the transmitted sub-stream $\mathbf{b}_{s}$, leading to a uniform error $e_{b,s}$  across all bit positions within pixels of the $s$-th segment.  By denoting the BER as $\mathcal P^\mathrm{e}_{s}$, we have $
e_{b,s} =\mathcal P^\mathrm{e}_{s},  ~ b=1,\dots, B$. The IMSE in \eqref{eq11} can be yielded as:
\begin{equation}\label{eq:imse2}
\mathrm{imse}(p_{s})  
=\sum_{s=1}^{S}\gamma_{s} \frac{4^{B}-1}{3}\alpha\exp\left(\beta\frac{p_{s}\vert h_{s} \vert^2}{\sigma^2}\right),
\end{equation}where $p_s$ is the power allocated to each symbol of the $s$-th sub-stream over the sub-channel $h_s$. 

In the case of SP-SS-I partitioning, $\mathbf{B}_b^{(s)}$ comprises the bits that form the transmitted sub-stream $\mathbf{b}_{b,s}$.  By denoting the BER as $\mathcal P^\mathrm{e}_{b,s}$, we have $
	e_{b,s} =\mathcal P^\mathrm{e}_{b,s}$, and the IMSE form in \eqref{eq11} can then be expressed as:
\begin{equation}\label{eq:imse3}
	\mathrm{imse}(p_{b,s}) 
	=\sum_{s=1}^{S} 	\sum_{b=1}^{B}\gamma_{s}\gamma_b \alpha\exp\left(\beta\frac{p_{b,s}\vert h_{b,s} \vert^2}{\sigma^2}\right),
\end{equation}where $p_{b,s}$ is the power allocated to each symbol of the $(b,s)$-th sub-stream over the sub-channel $h_{b,s}$. 	
}

{\color{sblue}{\textbf{Remark 1:}} It is important to note that {\bf Assumption 1} introduces a minor approximation to the MSE,	which is minimal in scenarios with infrequent communication errors and robust error-correcting mechanisms that effectively limit errors to at most a single bit per pixel. The assumption is particularly valid in high SNR regimes where the probability of multiple bit errors becomes statistically negligible. This approximation subsequently affects the IMSE calculation presented in \eqref{eq:imse1}, \eqref{eq:imse2}, and \eqref{eq:imse3}. Additionally, any potential mismatch between the BER model in  \eqref{eq:BER_SNR} and actual transmission conditions may result in minor deviation in  the IMSE formulation. }

\section{Data-Importance-Aware Waterfilling for Optimal Power Allocation \label{sec4}}
In this section,  power allocation problems are formulated within the proposed data-importance-aware communication framework. The objective is to minimize the task-oriented IMSE  subject to  total power constraints. To solve these problems, importance-aware waterfilling methods are developed, yielding optimal power allocation strategies. These approaches provide novel insights into  power resource prioritization that accounts for both data importance and channel conditions.  
 
\subsection{Optimal Power Allocation with SP-I Partitioning}

\begin{figure}[tp]
	\vspace{1em}
	\centering
	\includegraphics[width=0.95\columnwidth]{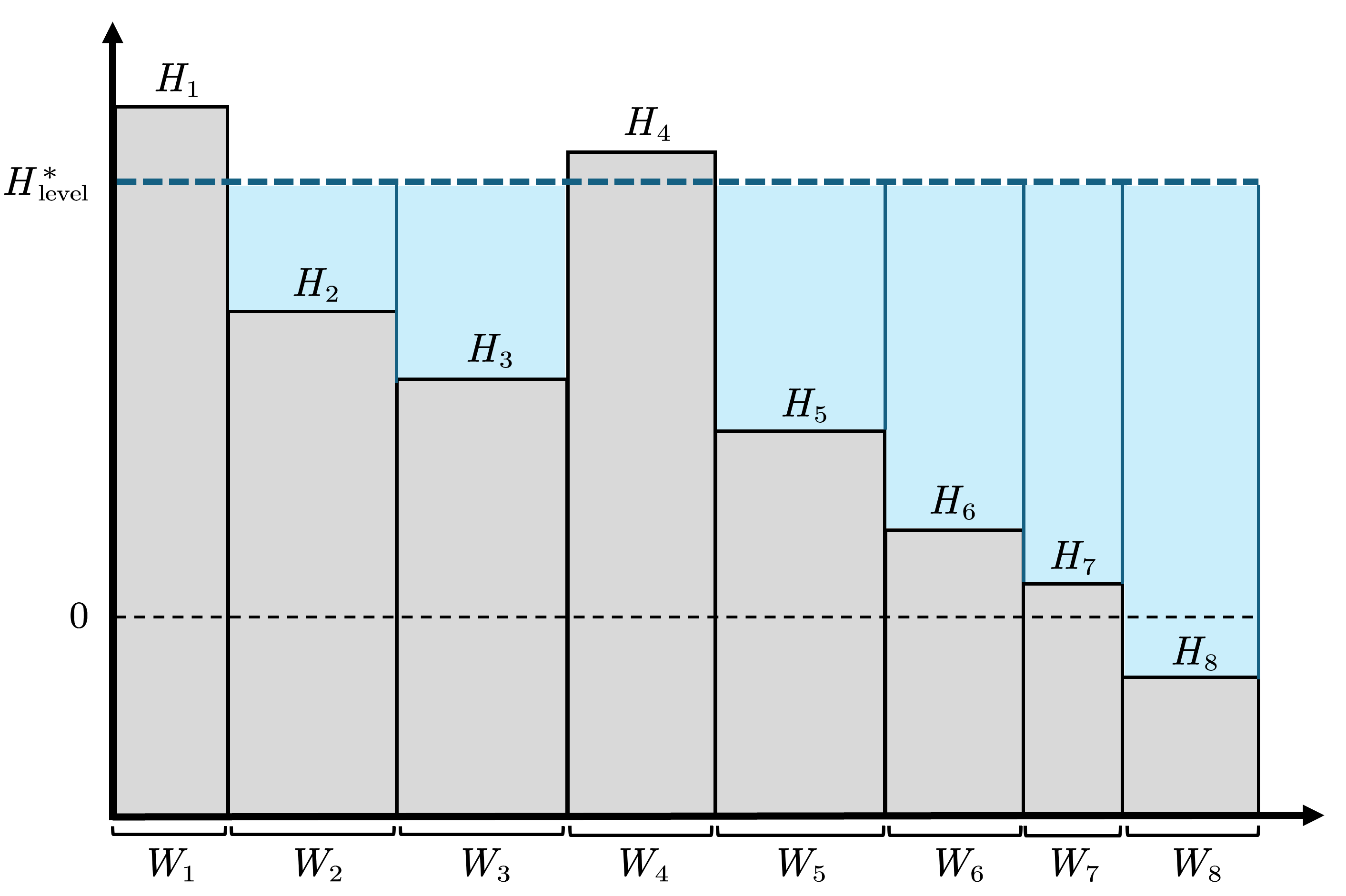}
	
	\caption{ An illustration of data-importance-aware waterfilling solution with SP-I partitioning, where $B=8$.}
	\label{fig:Waterfilling-BP-I}
\end{figure}

With SP-I partitioning, the bit length of each sub-stream $\mathbf{b}_b$ equals $I$,  {\color{sblue} and the number of modulated symbols is denoted as $L_b=\frac{I}{R\log_2 M}$.} The power allocation problem, which minimizes the IMSE {\color{sblue} in \eqref{eq:imse1} } subject to the total power constraint, is formulated as:


\begin{subequations}\label{eq:Prob_bit_level}
	\begin{align}
		\mathllap{(\mathcal P1)}\quad  \min_{p_b} \quad & \mathrm{imse}(p_b)\\
		\mathrm{s.t.} \quad &  {\color{sblue}L_b}p_b \le P, \label{eq:cons_power}
	\end{align}
\end{subequations}where $P$ is the total power budget.

Problem $(\mathcal P1)$ is convex with respect to (w.r.t.) the allocated power $p_b$ due to the convexity of the BER function in (\ref{eq:BER_SNR}).  Thereby, the optimal solution exists and can be obtained via the Lagrange multiplier technique \cite{boyd2004convex}.  The  corresponding Lagrange function is given by: 	
\begin{equation}
	\mathcal L(p_b, \lambda) \triangleq \mathrm{imse}(p_b) + \lambda\left( \sum_{b=1}^{B} {\color{sblue}L_b}p_b -P\right),
\end{equation}where $\lambda$ is the Lagrange multiplier. According to the Karush-Kuhn-Tucker (KKT) condition, the optimal solution satisfies:
\begin{equation}
		\frac{\partial \mathcal L(p_b, \lambda)}{\partial p_b}=\gamma_b\frac{\partial \mathcal P_{b}^{\mathrm{e}}}{\partial p_b} + {\color{sblue}L_b}\lambda=0,
	\end{equation}where $
	\frac{\partial \mathcal P_{b}^{\mathrm{e}}}{\partial p_b} = \alpha\beta\frac{\vert h_b \vert^2}{\sigma^2}\exp\left(\beta\frac{p_b\vert h_b \vert^2}{\sigma^2}\right)$.
	
	Since the allocated power cannot be negative, the optimal solution $p_b^*$ is derived as:

\begin{align}\label{eq:allop_bit_level}
		p_b^* & = \left( \frac{\sigma^2}{\beta \vert h_b \vert^2}\ln \frac{-\sigma^2{\color{sblue}L_b}\lambda^*}{\alpha\beta \gamma_b\vert h_b \vert^2} \right)^+\nonumber\\
		& = \underbrace{-\frac{\sigma^2}{\beta \vert h_b \vert^2}}_{W_b}\left(\underbrace{\ln \frac{-\alpha\beta}{\lambda^*} }_{H^*_\mathrm{level}} - \underbrace{\ln\frac{{\color{sblue}L_b}\sigma^2}{\gamma_b\vert h_b \vert^2}}_{H_b} \right)^+,
\end{align}where $(\cdot)^+$ denotes the  $\max(0,\cdot)$ operation, and $\lambda^*$ is the optimal Lagrange multiplier solution to the dual problem of  ($\mathcal P1$). This forms the waterfilling solution as illustrated in Fig. \ref{fig:Waterfilling-BP-I} with $B=8$, where $W_{b}$ and $H_{b}$ can be interpreted as the base widths and heights, respectively. $H_\mathrm{level}^*$ represents the optimal water level corresponding to the optimal $\lambda^*$. It satisfies the equality of power constraint (\ref{eq:cons_power}): 
\begin{equation}
\sum_{b=1}^{B} {\color{sblue}L_b}{W_b}\left({H^*_\mathrm{level}} - {H_b} \right)^+ = P,
\end{equation} {\color{sblue}which can be optimally solved. The procedure of determining the optimal $p_b^*$ is summarized in \textbf{Algorithm \ref{alg:alg1}}. The required computational complexity  is $\mathcal O(B\log_2 \delta)$, where $\delta$ is the tolerance threshold. Both computation time and memory requirements  scale  linearly with $B$, demonstrating well scalability characteristics that make it particularly well-suited for practical real-time communication systems where low latency is crucial.}

 \begin{algorithm}[t] 
	\renewcommand{\algorithmicrequire}{\textbf{Input:}}
	\renewcommand{\algorithmicensure}{\textbf{Output:}}
	\caption{Data-Importance-Aware Waterfilling with SP-I Partitioning.} 
	\label{alg:alg1}
	\begin{algorithmic}[1]		
		
		\STATE  Initialize water level $H_{\mathrm{level}}$:
		\[H_{\mathrm{level}} = \frac{ P+\sum_{b=1}^{B}{\color{sblue}L_b}W_b H_b}{\sum_{b=1}^{B}{\color{sblue}L_b}W_b}\]
		\STATE  Initialize $p_b$ based on (\ref{eq:allop_bit_level})
		\STATE \textbf{While} $\vert \sum_{b=1}^{B}{\color{sblue}L_b}p_b-P\vert/P\ge \delta$
		\STATE  \quad Update water level $H_{\mathrm{level}}$:
		\[H_{\mathrm{level}} \leftarrow H_{\mathrm{level}} - \frac{\sum_{b=1}^{B}{\color{sblue}L_b}p_b-P}{\sum_{b=1}^{B} {\color{sblue}L_b}W_b}\]\\
		 \STATE  \quad Compute $p_b$ based on (\ref{eq:allop_bit_level})		
		\STATE  \textbf{End}\\
		\STATE  {\color{sblue} Obtain and output $p_b^*$}
	\end{algorithmic}  
		
\end{algorithm}

\subsection{Optimal Power Allocation with  SS-I Partitioning}

 
With  SS-I partitioning, the bit length of the $s$-th sub-stream $\mathbf{b}_s$ equals $I_sB$, where $I_s$ is the number of pixels within the $s$-th segment. The power allocation problem, which minimizes the IMSE {\color{sblue}in \eqref{eq:imse2} subject to} the total power constraint, is formulated as:
\begin{subequations}\label{eq:Prob_segment_level}
	\begin{align}
		\mathllap{(\mathcal P2)}\quad \min_{p_s} \quad & \mathrm{imse}(p_s)\\
		\mathrm{s.t.} \quad & \sum_{s=1}^{S} {\color{sblue}L_s}p_s \le P,\label{eq:cons_power2}
	\end{align}
\end{subequations}{\color{sblue}where  $L_s = \frac{I_sB}{R\log_2 M}$ represents the number of symbols of the $s$-th sub-stream. }

Problem $(\mathcal P2)$ is convex w.r.t. $p_s$, which can be optimally solved using the Lagrange multiplier technique. Similarly, the Lagrange function by introducing the Lagrange multiplier $\lambda$ is given by:

\begin{equation}
\mathcal L(p_s, \lambda) \triangleq \mathrm{imse}(p_s) + \lambda\left(\sum_{s=1}^{S} {\color{sblue}L_s} p_s -P\right).
\end{equation}The optimal solution according to the KKT condition satisfies: 
\begin{equation}
\frac{\partial \mathcal L(p_s, \lambda)}{\partial p_s}=\gamma_{s} \frac{ 4^B-1}{3 } \frac{\partial \mathcal P_{s}^{\mathrm{e}}}{\partial p_s}+ {\color{sblue}L_s}\lambda=0,
\end{equation}where $
\frac{\partial \mathcal P_{s}^{\mathrm{e}}}{\partial p_s} = \alpha\beta\frac{\vert h_s \vert^2}{\sigma^2}\exp\left(\beta\frac{p_s\vert h_s \vert^2}{\sigma^2}\right)$. 

\begin{figure}[tp]
	\vspace{1em}
	\centering
	\includegraphics[width=0.95\columnwidth]{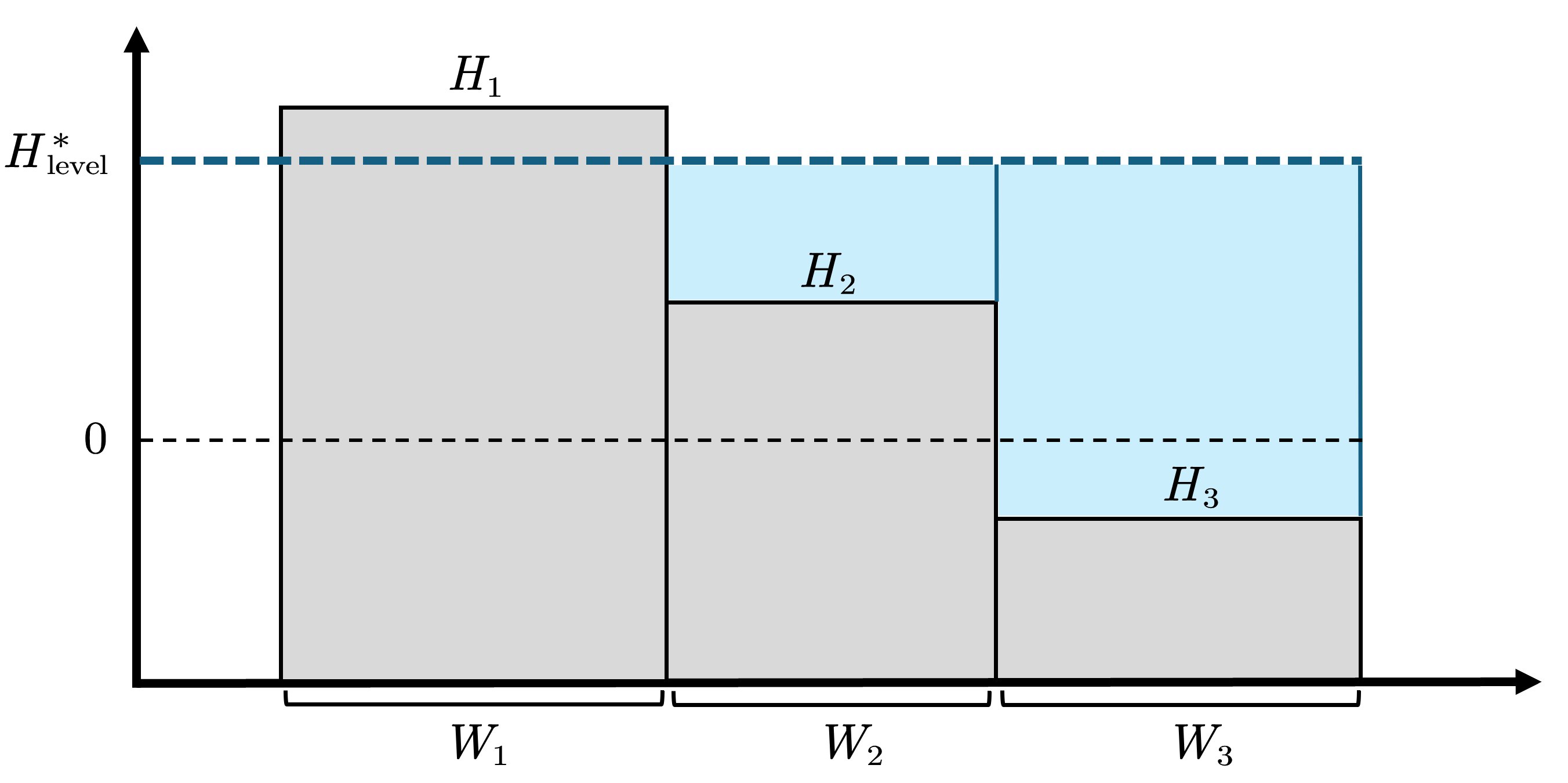}
	
	\caption{ An illustration of importance-aware waterfilling solution with SS-I partitioning, where $S=3$.}
	\label{fig:Waterfilling-SS-I}
\end{figure}

Since the allocated power cannot be negative, the optimal solution $p_b^*$ is derived as

\begin{align}\label{eq:allop_segment_level}
	p_s^* & = \left( \frac{\sigma^2}{\beta \vert h_s \vert^2}\ln \frac{-3\sigma^2{\color{sblue}L_s}\lambda^*}{\alpha\beta \left(4^B-1\right) \gamma_s \vert h_s\vert^2} \right)^+\nonumber\\
	& =  \underbrace{\frac{-\sigma^2}{\beta \vert h_s \vert^2}}_{W_s}\left(\underbrace{\ln\frac{-\alpha\beta \left(4^B-1\right)}{3\lambda^*}}_{H^*_\mathrm{level}} - \underbrace{\ln \frac{{\color{sblue}L_s}\sigma^2}{\gamma_s \vert h_s \vert^2} }_{H_s} \right)^+,
\end{align}where $\lambda^*$ is the optimal Lagrange multiplier solution to the dual problem of $(\mathcal P2)$.  Fig. \ref{fig:Waterfilling-SS-I} gives an illustration of the data-importance-aware waterfilling solution with the SS-I partitioning.  $W_s$, $H_\mathrm{level}^*$, and $H_s$ represent the base widths, optimal water level, and  base heights respectively. 
 The optimal power level $H_\mathrm{level}^*$, corresponding to $\lambda^*$, satisfies the equality of  power constraint (\ref{eq:cons_power2}):
\begin{equation}
\sum_{s=1}^{S} {\color{sblue}L_s}W_s \left(H^*_\mathrm{level} - H_s\right)^+ = P,
\end{equation}{\color{sblue}which can be optimally solved. The procedure of solving the optimal $p_s^*$ is summarized in \textbf{Algorithm \ref{alg:alg2}}.  This algorithm has the required computational complexity of $\mathcal O(S\log_2 \delta)$, with memory usage that scales linearly with the number of semantic segments $S$, making it efficient for real-time communication applications.}

\subsection{Optimal Power Allocation with SP-SS-I Partitioning}

With  SP-SS-I partitioning, the bit length of the $(b,s)$-th sub-stream $\mathbf{b}_{b,s}$ equals the number of pixels in the $s$-th segment $I_s$. The power allocation problem, which minimizes the IMSE {\color{sblue}in \eqref{eq:imse3}} under the total power constraint, is formulated as:
\begin{subequations}\label{eq:Prob_segment_bit_level}
	\begin{align}
		\mathllap{(\mathcal P3)}\quad \min_{p_{b,s}} \quad & \mathrm{imse}(p_{b,s})\\
		\mathrm{s.t.} \quad & \sum_{s=1}^{S}\sum_{b=1}^{B} {\color{sblue}L_{b,s}} p_{b,s} \le P, \label{eq:cons_power3}
	\end{align}
\end{subequations}{\color{sblue}where $L_{b,s} = \frac{I_s}{R\log_2 M}$ represents the number of modulated symbols of the $(b, s)$-th sub-stream. }

Problem $(\mathcal P3)$ is convex w.r.t. $p_{b,s}$, which can be optimally solved using the Lagrange multiplier technique. Similarly, the Lagrange function is given by:
\begin{equation}
	\mathcal L(p_{b,s}, \lambda)\triangleq \mathrm{imse}(p_{b,s}) + \lambda\left( \sum_{s=1}^{S}\sum_{b=1}^{B} {\color{sblue}L_{b,s}}p_{b,s} -P\right).
\end{equation}The optimal solution according to the KKT condition satisfies: 
\begin{equation}
	\frac{\partial \mathcal L(p_{b,s}, \lambda)}{\partial p_{b,s}}= \gamma_b \gamma_{s}\frac{\partial \mathcal P_{b,s}^{\mathrm{e}}}{\partial p_{b,s}}+ {\color{sblue}L_{b,s}}\lambda=0,
\end{equation}where $\frac{\partial \mathcal P_{b,s}^{\mathrm{e}}}{\partial p_{b,s}} = \alpha\beta\frac{\vert h_{b,s} \vert^2}{\sigma^2}\exp\left(\beta\frac{p_{b,s}\vert h_{b,s} \vert^2}{\sigma^2}\right)$.

\begin{figure}[tbp]
	\vspace{1em}
	\centering
	\includegraphics[width=0.95\columnwidth]{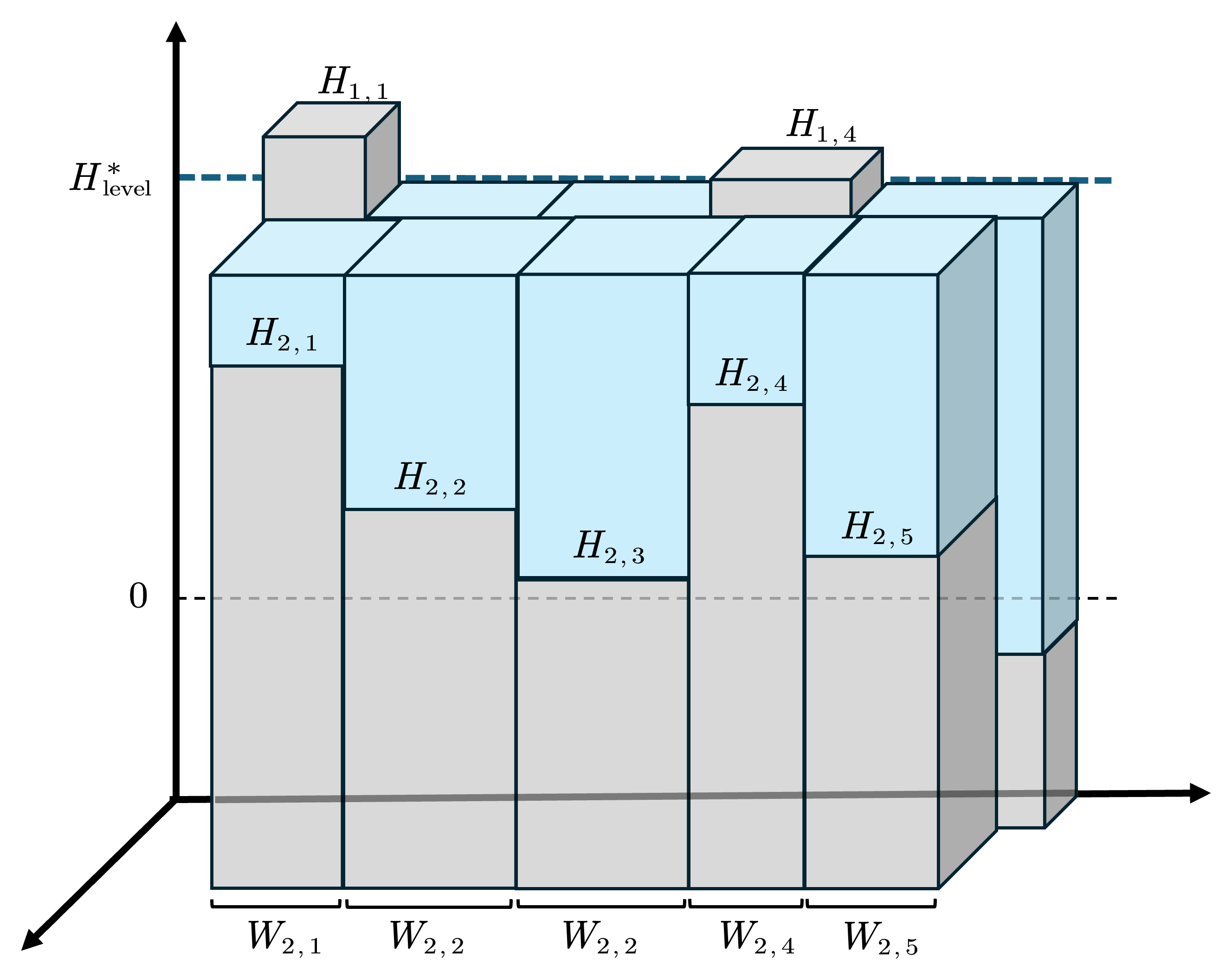}
	
	\caption{An illustration of data-importance-aware waterfilling solution with SP-SS-I partitioning, where $B=5$, $S=2$.}
	\label{fig:Waterfilling-C-BPSS-I}
\end{figure}

Since the allocated power cannot be negative, the optimal solution $p_{b,s}^*$ is derived as
	\begin{align}\label{eq:allop_segment_bit_level}
		 p_{b,s}^*& = \left( \frac{\sigma^2}{\beta \vert h_{b,s} \vert^2}\ln \frac{-\sigma^2 {\color{sblue}L_{b,s}}\lambda^*}{\alpha\beta  \gamma_b\gamma_s \vert h_{b,s} \vert^2} \right)^+\nonumber \\
		& =  \underbrace{\frac{-\sigma^2}{\beta \vert h_{b,s} \vert^2}}_{W_{b,s}}\left(\underbrace{\ln \frac{-\alpha\beta }{\lambda^*} }_{H^*_\mathrm{level}} -\underbrace{ \ln \frac{{\color{sblue}L_{b,s}}\sigma^2}{\gamma_s\gamma_b\vert h_{b,s} \vert^2} }_{H_{b,s}} \right)^+,
	\end{align}where  $\lambda^*$ is the optimal Lagrange multiplier solution to the dual problem of $(\mathcal P3)$.  This forms the three dimensional   waterfilling solution as illustrated in Fig. \ref{fig:Waterfilling-C-BPSS-I}, where $B=5$ and $S=2$. The terms $W_{b,s}$ and $H_{b,s}$ represent the base widths and heights, respectively.  The optimal power level $H_\mathrm{level}^*$, corresponding to $\lambda^*$, satisfies the equality of  power constraint (\ref{eq:cons_power3}):

\begin{equation}
	\sum_{s=1}^{S}\sum_{b=1}^{B} {\color{sblue}L_{b,s}} W_{b,s} \left(H^*_\mathrm{level} - H_{b,s}\right)^+ = P,
\end{equation}{\color{sblue}which can be optimally solved. The procedure of solving the optimal $p_{b,s}^*$ is summarized in \textbf{Algorithm \ref{alg:alg3}}. The computational requirements of this algorithm are characterized by complexity $\mathcal O(SB\log_2 \delta)$, with memory usage that scales linearly with the number of  sub-streams $SB$, ensuring its suitability for real-time communication scenarios.}

\begin{algorithm}[t]
	\renewcommand{\algorithmicrequire}{\textbf{Input:}}
	\renewcommand{\algorithmicensure}{\textbf{Output:}}
	\caption{ Data-Importance-Aware Waterfilling with SS-I Partitioning.} 
	\label{alg:alg2}
	
	\begin{algorithmic}[1]		
		
		\STATE  Initialize water level $H_{\mathrm{level}}$:\\
		\[H_{\mathrm{level}} = \frac{ P+\sum_{s=1}^{S}{\color{sblue}L_s}W_s H_s}{\sum_{s=1}^{S}{\color{sblue}L_s}W_s}\]
		\STATE Initialize $p_s$ based on (\ref{eq:allop_segment_level})
		
		\STATE \textbf{While} $\vert\sum_{s=1}^{S}{\color{sblue}L_s}p_s-P\vert/P\ge \delta$
		
		\STATE \quad Update water level $H_{\mathrm{level}}$:\\
		\[H_{\mathrm{level}} \leftarrow H_{\mathrm{level}} - \frac{\sum_{s=1}^{S}{\color{sblue}L_s}p_s-P} {\sum_{s=1}^{S}{\color{sblue}L_s}W_s}
		\]\\
		
		\STATE \quad Compute $p_k$ based on (\ref{eq:allop_segment_level})
		
		\STATE  \textbf{End}\\
		\STATE  {\color{sblue} Obtain and output $p_s^*$}
	\end{algorithmic}  
\end{algorithm}

\begin{algorithm}[t]
	\renewcommand{\algorithmicrequire}{\textbf{Input:}}
	\renewcommand{\algorithmicensure}{\textbf{Output:}}
	\caption{Data-Importance-Aware Waterfilling with  SP-SS-I Partitioning.} 
	\label{alg:alg3}
	
	\begin{algorithmic}[1]		
		
		\STATE  Initialize water level $H_{\mathrm{level}}$:\\
		\[H_{\mathrm{level}} = \frac{ P+\sum_{s=1}^{S}\sum_{b=1}^{B} {\color{sblue}L_{b,s}}W_{b,s} H_{b,s}}{\sum_{s=1}^{S}\sum_{b=1}^{B}{\color{sblue}L_{b,s}}W_{b,s}}\]
		\STATE Initialize $p_{b,s}$ based on (\ref{eq:allop_segment_bit_level})
		
		\STATE \textbf{While} {\color{sblue}{$\vert\sum_{s=1}^{S}\sum_{b=1}^{B}{\color{sblue}L_{b,s}}p_{b,s}-P\vert/P\ge \delta$}}
		
		\STATE \quad Update water level $H_{\mathrm{level}}$:\\
		\[H_{\mathrm{level}} \leftarrow H_{\mathrm{level}} - \frac{\sum_{s=1}^{S}\sum_{b=1}^{B}{\color{sblue}L_{b,s}}p_{b,s}-P} {\sum_{s=1}^{S}\sum_{b=1}^{B}{\color{sblue}L_{b,s}}W_{b,s}}
		\]\\
		
		\STATE \quad Compute $p_{b,s}$ based on (\ref{eq:allop_segment_bit_level})
		
		\STATE  \textbf{End}\\
		\STATE  {\color{sblue} Obtain and output $p_{b,s}^*$}
	\end{algorithmic}  
\end{algorithm}

\subsection{Waterfilling Gain and Novel Insights}
In this section, three data-importance-aware waterfilling methods have been developed to optimally allocate power across sub-streams under the proposed SP-I, SS-I, and SP-SS-I partitioning criteria. {\color{sblue}For a system with $K$
sub-streams, each characterized by importance weight $\omega_k$ and symbol length $L_k$, the explicit relationship between IMSE and signal power can be expressed in a unified form as: }
\begin{equation}\label{eq:imse_k}
	\mathrm{imse}(p_k) = \sum_{k=1}^K\omega_k \alpha \exp\left(\beta\frac{p_{k}\vert h_{k} \vert^2}{\sigma^2}\right).
\end{equation}Here, $\omega_k$ is equal to $\gamma_b$, $\gamma_s\frac{4^{B}-1}{3}$ and $\gamma_b\gamma_s$ under the SP-I, SS-I and SP-SS-I partitioning criteria, respectively. The optimal solution of the $k$-th sub-stream $p_k^*$ is expressed by: 
\begin{equation}\label{eq:p_k_solution}
	p_k^* =  \underbrace{\frac{-\sigma^2}{\beta \vert h_{k} \vert^2}}_{W_{k}}\left(\underbrace{\ln \frac{-\alpha\beta }{\lambda^*} }_{H^*_\mathrm{level}} -\underbrace{ \ln \frac{{\color{sblue}L_{k}}\sigma^2}{\omega_k\vert h_{k} \vert^2} }_{H_{k}} \right)^+.\\	
\end{equation} The optimal water level  $H_\mathrm{level}^*,$ satisfying the equality of the power constraint, is determined by BER parameters $\alpha$ and $\beta$, and the optimal Lagrange multiplier $\lambda^*$. For the $k$-th sub-stream,  $W_k$ is jointly determined by the channel condition $\vert h_k\vert^2/\sigma^2$, and the BER parameter $\beta$.  $H_k$ is jointly determined by the {\color{sblue}symbol length $L_k$}, channel condition $\vert h_k\vert^2/\sigma^2$,  and importance wights $\omega_k$. Substituting \eqref{eq:p_k_solution} back to \eqref{eq:imse_k} yields the optimal IMSE, given by {\color{sblue} $\mathrm{imse}^*= \sum_{k=1}^K \omega_k\alpha\mathcal\exp\big( -\big(H_\mathrm{level}^* - \ln \frac{{\color{sblue}L_{k}}\sigma^2}{\omega_k\vert h_{k} \vert^2}\big)^+\big)$ }

To measure the data-importance-aware waterfilling gain, we adopt the conventional waterfilling method as the benchmark, which accounts for the channel but treat all sub-streams equally. Since the transmission rate is determined by the adopted channel coding rate and modulation order, the MA waterfilling method \cite{ma2008bit,jinPower2022}  is adopted. The objective is to minimize the sum MSE subject to the power constraint, formulating the problem as:
\begin{equation}\label{eq:Prob_MA}
	\min_{p_k} \sum_{k=1}^K \alpha \exp\left(\beta\frac{p_{k}\vert h_{k} \vert^2}{\sigma^2}\right) ~
		\mathrm{s.t.}  \sum_{k=1}^{K} {\color{sblue}L_k} p_k \le P. 
\end{equation}
The solution to \eqref{eq:Prob_MA}  is given by
\begin{equation}\label{eq:MA_p}
	p^{*}_{\mathrm{MA},k} = \frac{-\sigma^2}{\beta \vert h_{k} \vert^2}\left(\underbrace{\ln \frac{-\alpha\beta}{\lambda^*}}_{H_\mathrm{MA,level}^*}  - \ln \frac{{\color{sblue}L_k}\sigma^2}{\vert h_{k} \vert^2} \right)^+. 
\end{equation}Substituting \eqref{eq:MA_p} back  to \eqref{eq:imse_k} yields {\color{sblue} $ 
	\mathrm{imse}^*_\mathrm{MA} = \sum_{k=1}^K \omega_k\alpha\mathcal\exp\big( -\big(H_\mathrm{MA,level}^* - \ln \frac{L_k\sigma^2}{\vert h_{k} \vert^2}\big)^+\big).$} The importance-aware waterfilling gain, denoted as $G_\mathrm{gain}$, is then defined as:
\begin{equation}\label{eq:gain}
	G_\mathrm{gain} = -10\log_{10}\left( \frac{\mathrm{imse}^*}{\mathrm{imse}^*_\mathrm{MA}}\right).
\end{equation}
This gain is more pronounced when the variation of importance weight $\omega_k$ exhibits greater variation  (as illustrated in Fig. \ref{fig:IMSE_Gains} in Sec. \ref{sec V:C}). This variation can be measured by Gini coefficient  \cite{hojeij2017waterfilling}, which is given by {\color{sblue}$
	 G_\mathrm{eff} = \frac{\sum_{i=1}^{K}\sum_{j=1}^{K}|\omega_i-\omega_j|}{2K\sum_{k=1}^{K}\omega_k}$.}

{\color{black}The above analysis reveals novel insights: 1) The optimal power allocation for each sub-stream depends on both its data importance and channel condition; 2) The optimal power of each sub-stream is monotonically increasing  with its importance weight, demonstrating that more critical sub-streams receive higher transmission priority; 3) 
The optimal power of each sub-stream initially increases with its channel gain up to a threshold. Beyond this point, the allocated power decreases, indicating that exceptionally good sub-channels are not necessarily prioritized; and 4) The data-importance-aware waterfilling gain becomes more pronounced when importance weights show greater variation.  }

\section{Simulation Results}

This section presents simulation results that demonstrate the effectiveness of the proposed data-importance-aware communication framework and validate the proposed data-importance-aware waterfilling methods in improving the task-oriented reconstruction performance. 
\begin{figure}[tp]
	\centering
	\includegraphics[width=1\columnwidth]{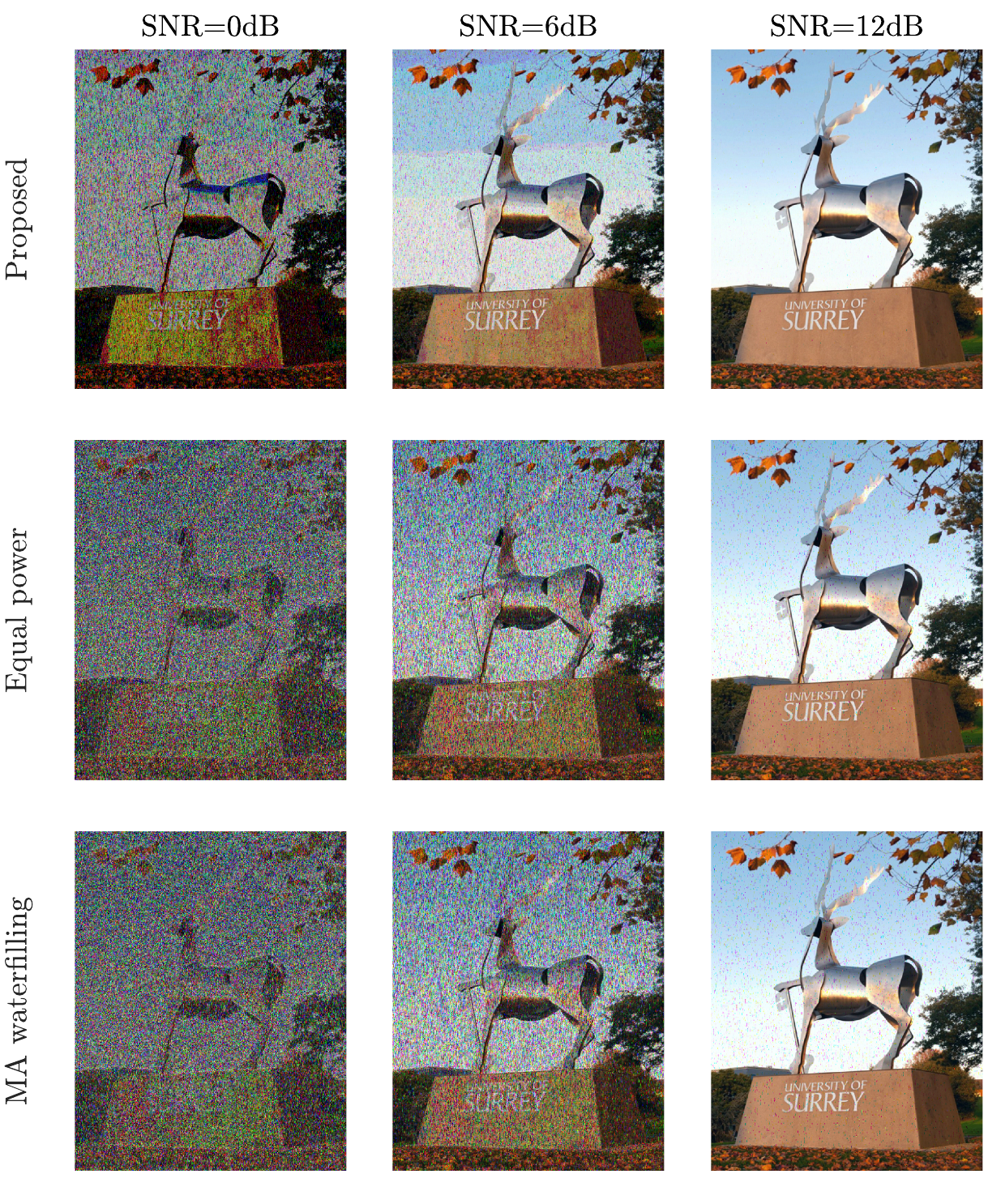}
	\caption{{\color{sblue}Visual quality of reconstructed images with  SP-I partitioning, where the bit importance weights are set to $\gamma_b=2^{2(b-1)}$.}}
	\label{fig:reconstructed_images_bit_levels}
\end{figure}
                           
\subsection{Parameter Setup}
The simulations consider a point-to-point data-importance-aware communication scenario using  random interleavers, convolutional codes and QAM modulations. Identical channel coding and modulation schemes are applied across all sub-streams, where the coding rate and the modulation order are set to $1/2$ and $M=16$, respectively. 
When employed, random interleaving is performed before channel coding and modulation. Under these settings,  the fitting parameters of BER function in (\ref{eq:BER_SNR}) are $\alpha=0.5123$ and $\beta=-0.2862$.  For the channels, Rayleigh fading with $h_k\sim\mathcal{CN}(0,1)$ is considered unless stated otherwise, and the noise variance is set to $\sigma^2=1$. 
 The source image is a $640\times 512$ RGB image with $B=8$ bits  per pixel per color channel. Using the state-of-the-art SAM, the image is segmented into three semantic regions: ``stag'', ``base'', and ``background''. These segments are assigned the importance weights of $\gamma_{\mathrm{stag}}=0.4975$, $\gamma_{\mathrm{base}}=0.4975$, and $\gamma_{\mathrm{background}}=0.0050$, respectively, unless stated otherwise, indicating that the ``stag'' and ``base'' are significantly  more important than the ``background''. {\color{sblue}It is important to note that these specific weight values are task-dependent and would vary based on the particular computer vision application at the receiver. Although we employ manually assigned weights for this proof-of-concept demonstration, the fundamental contributions of our work remain valid regardless of how these weights are determined.}   

For performance comparison, two baselines are considered: 1) Equal power allocation: The power is equally allocated to all modulated symbols, regardless of data importance and channel conditions; 2)  MA waterfilling: The allocated power is obtained  to  minimize  sum  BERs to adapt channel conditions regardless of data importance, which is given in \eqref{eq:MA_p}. {\color{sblue} The data-importance-aware communication framework and waterfilling approaches have potential to be integrated into real-world communication systems. The proposed framework is compatible with advanced coding schemes such as LDPC codes, requiring only updats to the BER function parameters $(\alpha, \beta)$ to obtain the optimal power allocation while leveraging enhanced error correction capabilities. In addition, the proposed power allocation approaches adapt efficiently to the dynamic channel conditions through periodic channel estimation and power allocation updates, with linear computational scaling that makes it well-suited for real-time system. We also acknowledge practical implementation challenges, primarily the determination of semantic importance weights and the need for semantic map transmission to ensure correct reconstruction at the receiver. }

\subsection{Visual {\color{sblue}Quality} of Reconstructed Images \label{sec:V_B}}

\begin{figure}[tbp]
	\centering
	\includegraphics[width=1\columnwidth]{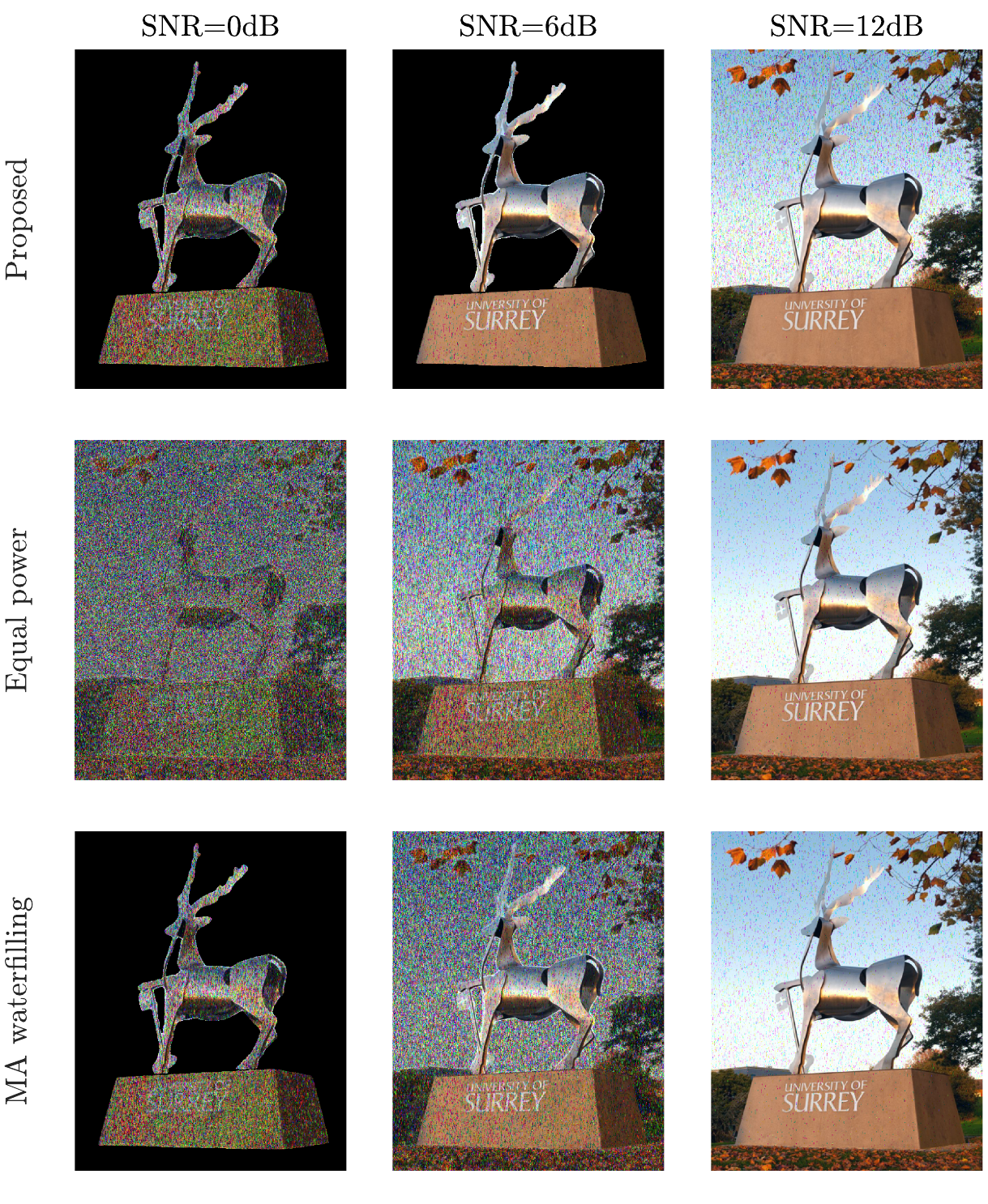}
	\caption{{\color{sblue}Visual quality of reconstructed images with  SS-I partitioning, where the semantic importance weights  are set to $\gamma_{\mathrm{stag}}=0.4975$, $\gamma_{\mathrm{base}}=0.4975$, and $\gamma_{\mathrm{background}}=0.0050$.}}
	\label{fig:reconstructed_images_segment_levels}
\end{figure}

\begin{figure}[thbp]
	\centering
	\includegraphics[width=1\columnwidth]{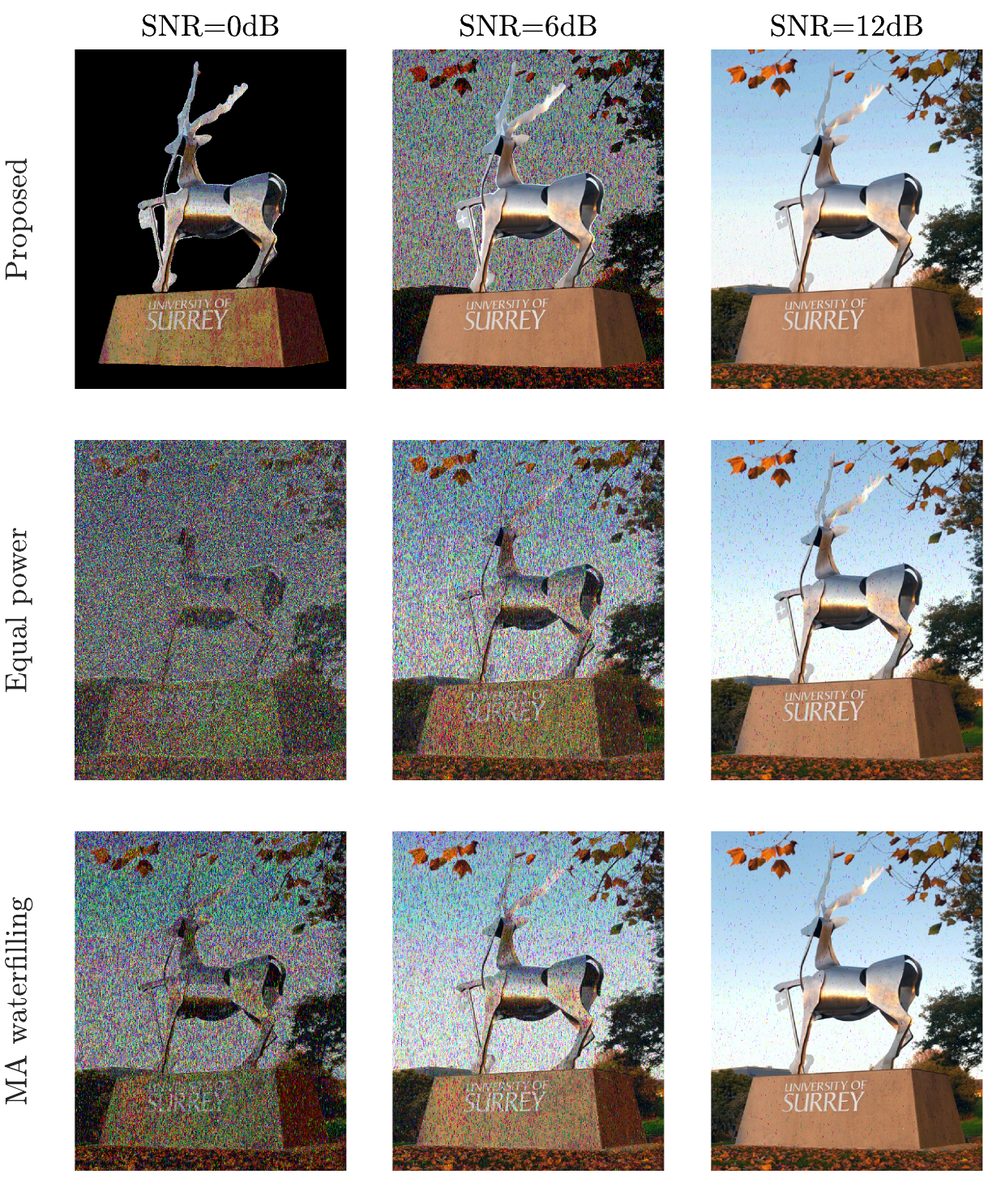}
	\caption{{\color{sblue} Visual quality of reconstructed images with  SP-SS-I partitioning, where bit importance weights are set to $\gamma_b=2^{2(b-1)}$, and semantic importance weights $\gamma_s$ are set to $\gamma_{\mathrm{stag}}=0.4975$, $\gamma_{\mathrm{base}}=0.4975$, and $\gamma_{\mathrm{background}}=0.0050$.}}
	\label{fig:reconstructed_images_segment_bit_levels}
\end{figure}

We examine the visual quality of the reconstructed images transmitted through AWGN channels without interleaving. Note that under AWGN channels, the power obtained using the MA waterfilling method is affected by the {\color{sblue}symbol lengths} of sub-streams according to \eqref{eq:MA_p}, {\color{sblue}reducing} to prioritize shorter sub-stream.  Figs. \ref{fig:reconstructed_images_bit_levels}, \ref{fig:reconstructed_images_segment_levels}, and \ref{fig:reconstructed_images_segment_bit_levels} compare the visual quality of reconstructed images with the proposed SP-I, SS-I, and SP-SS-I partitioning criteria, respectively. For each criteria,  the proposed data-importance-aware waterfilling approach is compared against equal power allocation and MA waterfilling methods.

\begin{figure}[thbp]
	\centering
	\includegraphics[width=1\columnwidth]{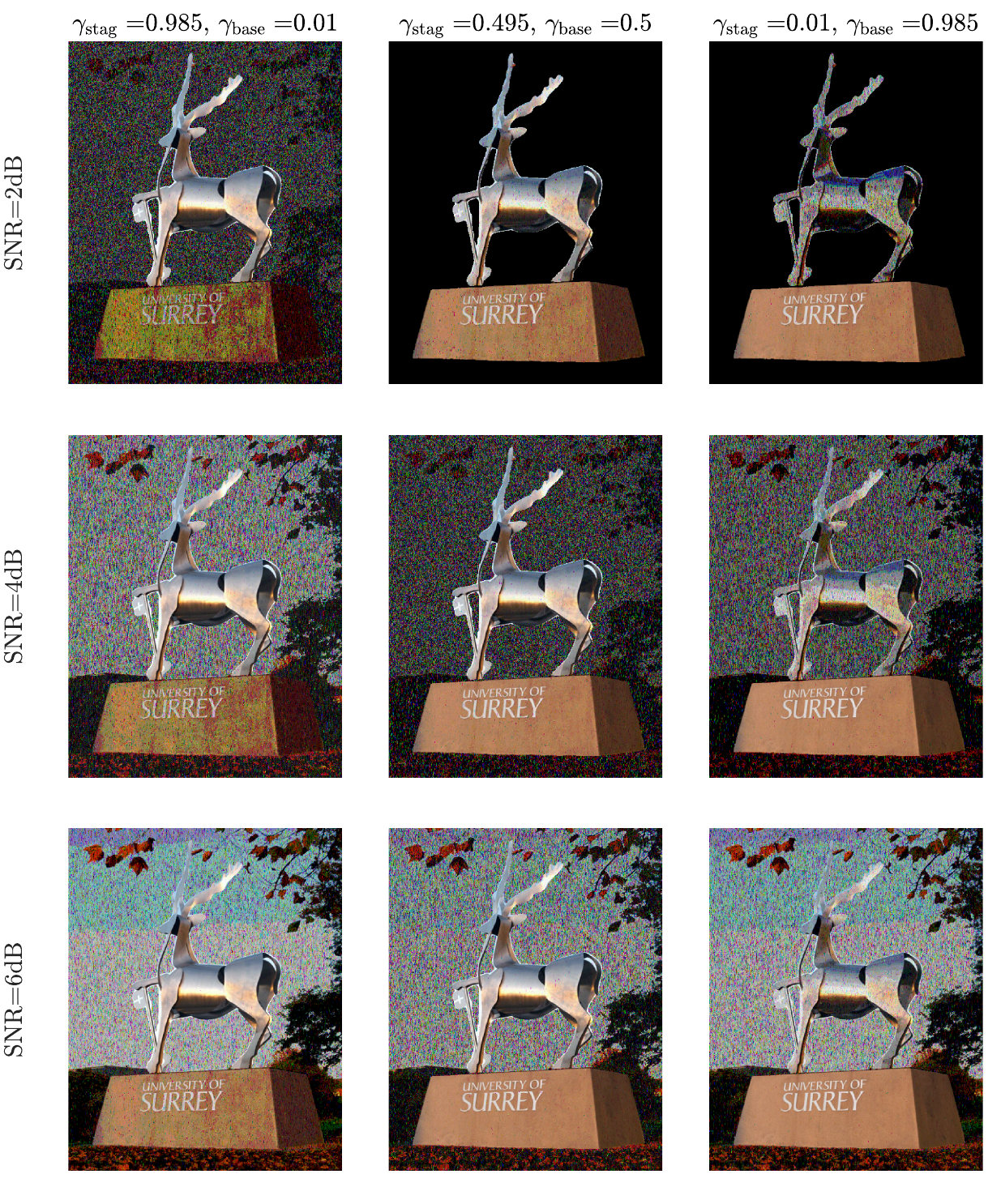}
	\caption{{\color{sblue}Visual quality of reconstructed images with  SP-SS-I partitioning, where bit importance weights are set to $\gamma_b=2^{2(b-1)}$ and semantic importance weights $\gamma_s$ are adjusted while maintaining a fixed weight of $\gamma_{\mathrm{background}}=0.0050$.}}
	\label{fig:reconstructed_images_segment_bit_levels_importance}
\end{figure}

With the SP-I partitioning (Fig. \ref{fig:reconstructed_images_bit_levels}), all segments can be considered to be assigned equal semantic importance weights.  As SNR increases, the visual quality improves across all methods, with our proposed approach demonstrating superior reconstruction quality. For example  at {\color{sblue}$\mathrm{SNR}=6\,$dB, key visual elements like ``University of Surrey'' and the ``stag'' are  recognizable  using the proposed approach, while these features remain indistinguishable with the baseline methods.} Note that the MA waterfilling becomes equivalent to the equal power allocation, and achieves identical visual quality, due to the unit channel gains and equal length of all sub-streams with the SP-I partitioning. 
For the SS-I and SP-SS-I partitioning (Figs. \ref{fig:reconstructed_images_segment_levels} and \ref{fig:reconstructed_images_segment_bit_levels}), recall that we assign the lowest importance weight ($0.0050$) to the ``background'' segment. Our proposed data-importance-aware approaches demonstrate superior visual quality in reconstructing the high-importance ``stag'' and ``base'' segments compared to MA waterfilling and equal power allocation methods. Notably, Fig. \ref{fig:reconstructed_images_segment_levels} shows that at low SNRs, power resources can be saved for high-importance data by excluding the least important segment from transmission.  Between these two baselines, the MA waterfilling method achieves better visual quality than equal power allocation, as it naturally penalizes longer sub-streams which in this case corresponds to the least important background segment.

Among the three importance-aware data partitioning criteria, the SP-SS-I partitioning achieves the best overall visual quality in image reconstruction. This superior performance stems from two factors.  Compared to the SP-I partitioning, it produces clearer reconstruction of the important ``stag'' and ``base'' segments by taking the segment importance into account. Compared to the SS-I model, it maintains better visual quality across all segments by incorporating varying levels of importance in bits within pixels.  To further analyze the impact of semantic weights $\gamma_s$, Fig. \ref{fig:reconstructed_images_segment_bit_levels_importance} depicts the  {\color{sblue}visual quality of the reconstructed images} with SP-SS-I partitioning, where the weights of ``stag'' and ``base'' segments are adjusted while  the ``background'' weight is fixed at $\gamma_\mathrm{background}=0.0050$. It demonstrates that segments with higher importance weights are reconstructed with  better visual qualities, as they receive larger power allocations under the proposed data-importance-aware waterfilling approach. {\color{sblue}These results also highlight the robustness of the proposed approaches to variations in semantic weights.}

\subsection{Reconstruction Performance Evaluation\label{sec V:C}}
\begin{figure}[tbp]
	\centering
	\includegraphics[width=0.95\columnwidth]{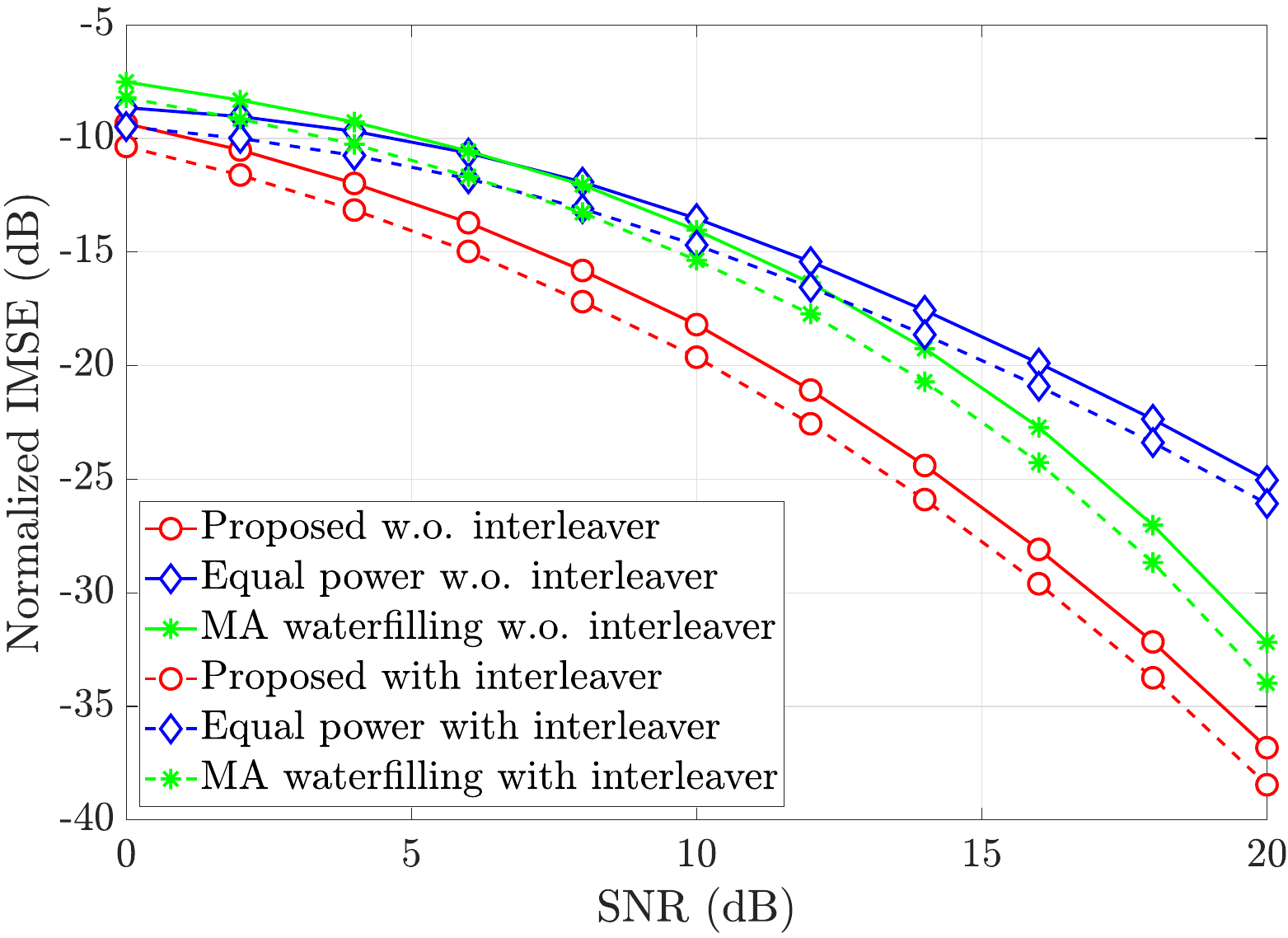}
	\caption{The normalized IMSE performance comparisons against the baselines using the SP-I partitioning. }
	\label{fig:MSE_VS_SNR_bit}
\end{figure}
\begin{figure}[t]
	\centering
	\includegraphics[width=0.95\columnwidth]{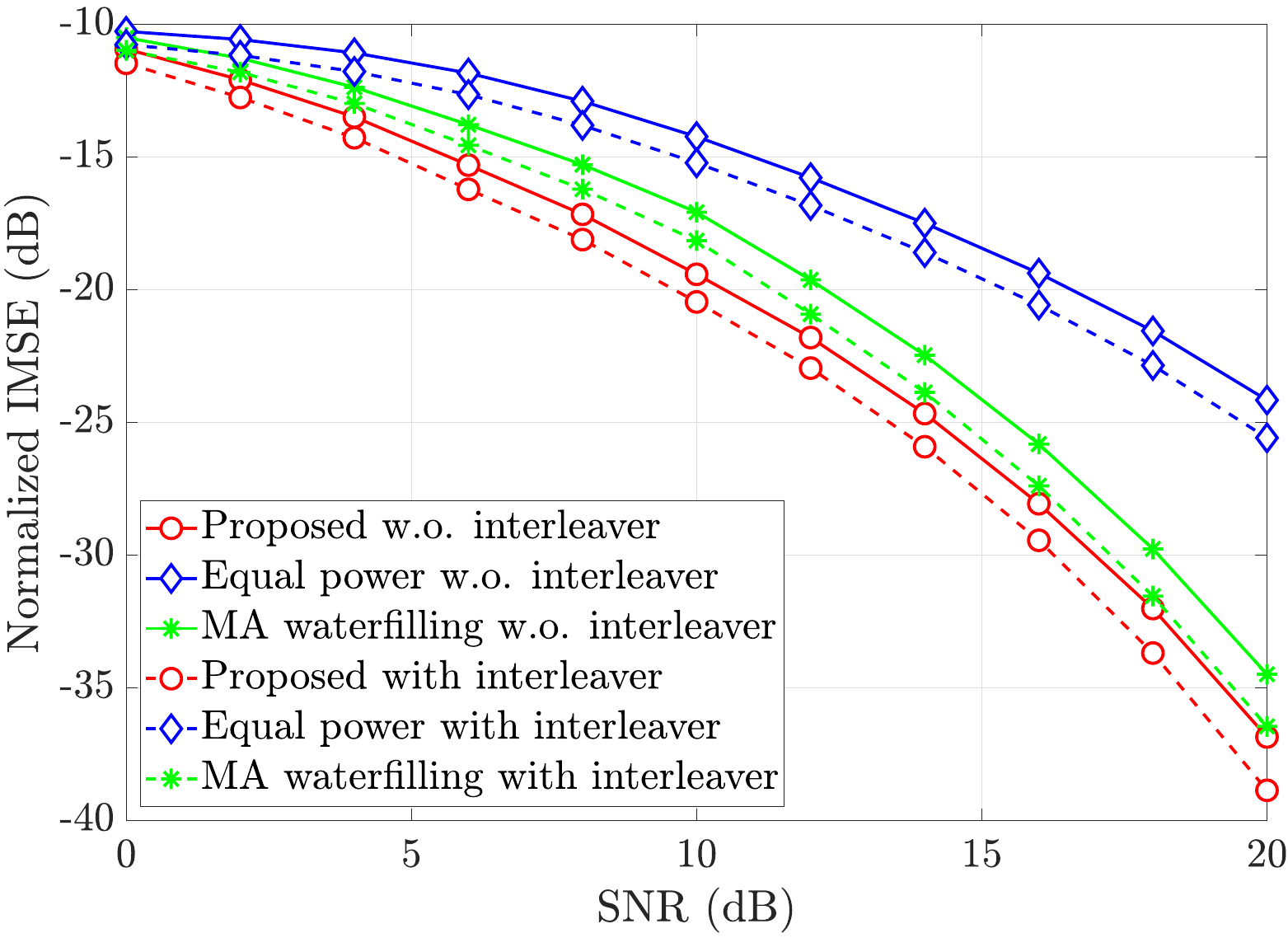}
	\caption{The normalized IMSE performance comparisons against the baselines using the SS-I partitioning.}
	\label{fig:MSE_VS_SNR_segment}
\end{figure}

\begin{figure}[t]
	\centering
	\includegraphics[width=0.95\columnwidth]{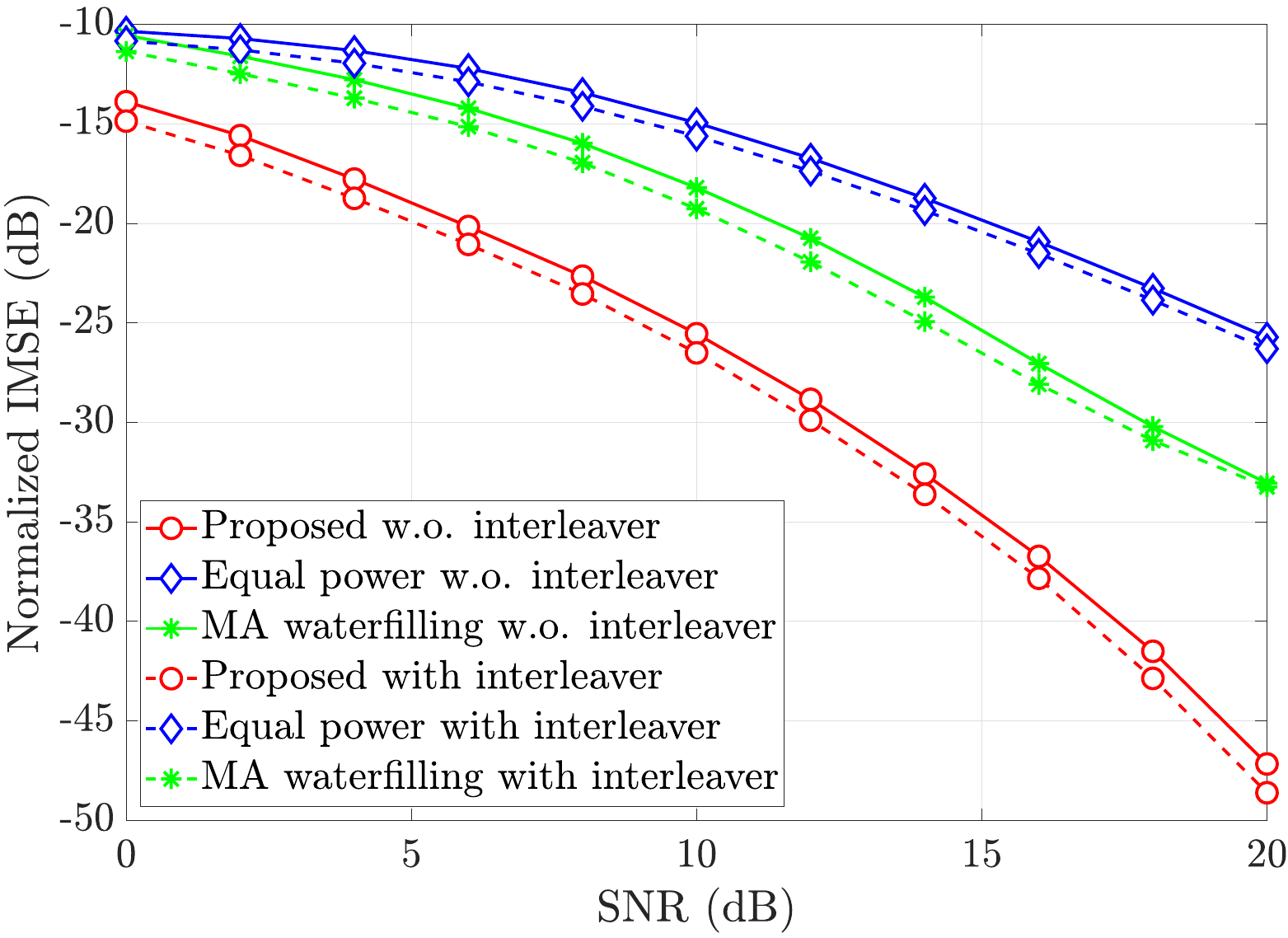}
	\caption{The normalized IMSE performance comparisons against the baselines using the SP-SS-I partitioning.}
	\label{fig:MSE_VS_SNR_segment_bit}
\end{figure}

\begin{figure}[t]
	\centering
	\includegraphics[width=0.95\columnwidth]{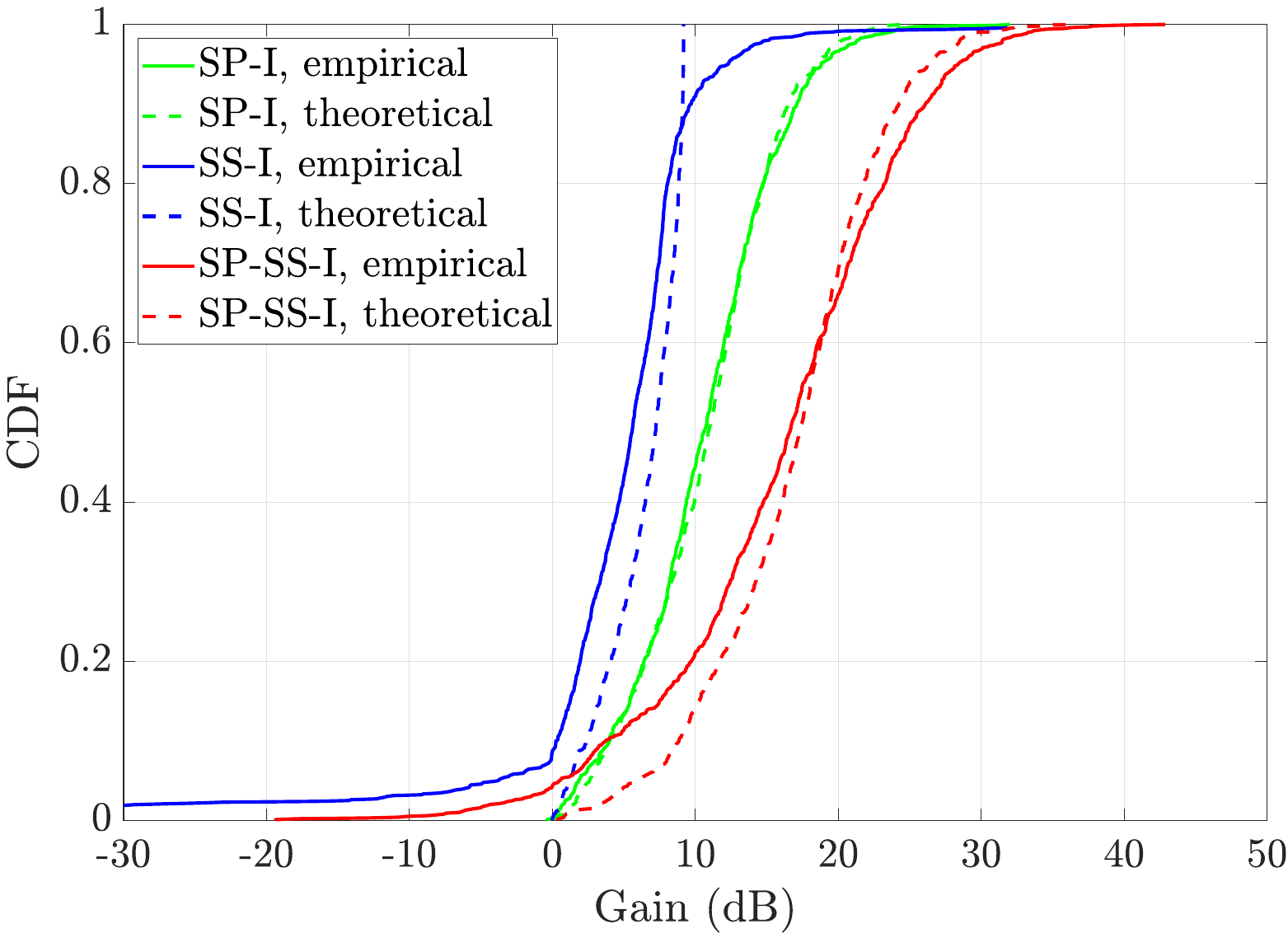}
	\caption{{\color{sblue}The data-importance-aware waterfilling gains under SP-I, SS-I and SP-SS-I partitioning at $\mathrm{SNR}=16\,$dB.}}
	\label{fig:IMSE_Gains}
\end{figure}

This subsection evaluates the task-oriented  reconstruction performance in terms of IMSEs normalized by $\Vert \mathbf{I}\Vert^2/I$. The results are averaged over 100 channel realizations both with and without implementations of random interleavers.  Figs. \ref{fig:MSE_VS_SNR_bit},  \ref{fig:MSE_VS_SNR_segment} and  \ref{fig:MSE_VS_SNR_segment_bit} compare the normalized IMSE performance of the proposed data-importance-aware waterfilling approaches against the baselines using SP-I, SS-I, and SP-SS-I partitioning criteria, respectively. Note that  the normalized IMSE is equivalent to the conventional normalized MSE with  SP-I partitioning, as the whole image can be considered as a segment. While the normalized IMSE decreases with SNR across all methods and data partitioning criteria, the proposed data-importance-aware approaches consistently achieve significantly lower normalized IMSE compared to the baselines. The MA waterfilling outperforms equal power allocation across all data partitioning criteria by adapting to channel conditions, {\color{sblue}unlike equal power allocation which distributes power uniformly regardless of channel quality. Additionally, random interleavers further improve the normalized IMSE performance by disrupting the correlation between sequential pixels, preventing  consecutive zeros and ones that weaken the error correction capability of convolutional codes.}

Figures \ref{fig:MSE_VS_SNR_bit}, \ref{fig:MSE_VS_SNR_segment}, and \ref{fig:MSE_VS_SNR_segment_bit} also demonstrate that the SP-SS-I partitioning achieves the most significant improvement in normalized IMSE when using the proposed data-importance-aware waterfilling approach, outperforming both SP-I and SS-I partitioning.  This superior performance stems from its dual-level partitioning strategy that considers both SP-I and SS-I models, with SP-I partitioning being the more substantial contributor. The strategy provides additional $S$ and $B$ degrees of freedom (DoF) compared to either SP-I or SS-I partitioning criteria alone {in optimizing power allocation based on data importance}.  The increased DoF enables greater variations in importance weights, as evidenced by the Gini coefficients: $0.7917$ for SP-I, $0.3283$ for SS-I, and $0.8574$ for SP-SS-I partitioning. At high SNRs ($\mathrm{SNR}>10\,$dB), the proposed approaches achieves normalized IMSE gains: more than $7\,$dB and $10\,$dB under the SP-SS-I partitioning, $4\,$dB and $4.5\,$dB under the SP-I partitioning, and $2.5\,$dB and $5.2\,$dB under the SS-I partitioning, compared to MA waterfilling and equal power allocation, respectively. To achieve a target normalized IMSE  of $-26\,\mathrm{dB}$, the proposed approach reduces the required SNR by $5\,\mathrm{dB}$ and $10\,\mathrm{dB}$  under the SP-SS-I partitioning, $3\,\mathrm{dB}$ and $6\,\mathrm{dB}$ under the SP-I partitioning, and $2\,\mathrm{dB}$ and $6\,\mathrm{dB}$ under the SS-I partitioning, compared to the baselines. These substantial performance improvements demonstrate the potential of the proposed data-importance-aware framework to enhance data efficiency and  robustness for real-time CV applications,  particularly in bandwidth-limited and resource-constrained environments.

Fig. \ref{fig:IMSE_Gains} depicts the empirical and theoretical CDFs of data-importance-aware gains (defined in \eqref{eq:gain}) with the SP-I, SS-I and SP-SS-I partitioning criteria at $\mathrm{SNR}=16\,$dB. The results demonstrate that the SP-SS-I partitioning, which exhibits the largest variation in importance weights, achieves the highest data-importance-aware gain, followed by SP-I and SS-I partitioning respectively. {\color{sblue} Theoretically, it is predicted that these gains are consistently  non-negative, indicating improved IMSE performance over the baseline approach. However, the empirical CDFs show small probabilities ($\le 0.1$) of negative gains across all three data partition criteria, with SS-I partition exhibiting the least favorable performance. This discrepancy stems from two factors. First,   \textbf{Assumption \ref{assp1}} made in our theoretical derivation of IMSE introduces approximation errors in scenarios with frequent communication errors. Specifically, when multiple bit errors occur within a single pixel, the actual performance deviates from the theoretical ones. This effect is particularly pronounced with SS-I partition criteria, where bits from the same semantic segment are grouped into the same data streams increasing the likelihood of clustered errors. Second,   the  mismatched between the fitting BER function and the actual transmission conditions further contributes to this deviation.}

\section{Conclusion}

This paper addressed two key challenges {\color{sblue} of developing task-oriented metrics and efficient power allocation strategies} in real-time CV  applications with critical latency constraints that preclude source coding.  The SP-I and SS-I model were first proposed to characterize data importance based on bit positions within pixels and semantic relevance within visual segments, leading to three importance-aware data partitioning criteria. A novel task-oriented metric, IMSE, was introduced to evaluate reconstructed images to capture both the task-specific significance of visual information and the interdependence between CV and communication performance. {\color{sblue}To minimize the IMSE, importance-aware waterfilling approaches were developed, yielding optimal power allocation strategies based on both data importance and channel conditions.} Simulation results  demonstrated the consistent superior performance of the proposed importance-aware waterfilling methods over equal power allocation and conventional waterfilling schemes in both visual reconstruction quality and IMSE across all data partitioning criteria. Among these, the SP-SS-I partitioning, with the greatest variations in importance weights, achieved the most signifiant performance improvements  compared to individual SP-I and SS-I criteria. 
These substantial performance improvements demonstrate potential of the proposed framework to improve data efficiency and robustness in real-time CV applications.


\begin{appendices}
\section{Fitting Results of the BER Function in \eqref{eq:BER_SNR} \label{appendixA}}
{\color{sblue}The paramaters $\alpha$ and $\beta$ of the BER function in \eqref{eq:BER_SNR} are obtained through data fitting. For the convolutional codes, we use poly2trellis(3, [6 7]) for the 1/2 rate and poly2trellis(3, [5 6 7]) for the 2/3 rate, with fitting parameters provided below.}
\begin{table}[th]
	\caption{The fitting parameters of the BER Function. 
    }
	\centering
	\begin{tabular}{|c|c|c|c|}
		\hline
		$(\alpha, \beta)$  & coding rate (1/2) & coding rate (2/3)\\
		\hline
		BPSK   & $(0.6559, -2.5484)$  & $(0.9774,-5.0670)$ \\
		\hline
		$4$-QAM   & $(0.5914, -1.1788)$  & $(0.7302, -2.1711)$ \\
		\hline
		$8$-QAM  & $(0.5271, -0.4368)$  & $(0.6256, -0.8691 )$ \\
		\hline
		$16$-QAM  & $(0.5123,-0.2862)$  & $(0.5699, -0.5604 )$ \\
		\hline
	\end{tabular}
\end{table}
 \end{appendices}

\bibliographystyle{IEEEtran}
\bibliography{reference}

\end{document}